\documentclass[showpacs,twocolumn]{revtex4}
\usepackage{amssymb}

\newcommand{\beq}{\begin{eqnarray}}
\newcommand{\eeq}{\end{eqnarray}}
\newcommand{\nneeq}{\nonumber \end{eqnarray}}
\newcommand{\nnn}{\nonumber}
\newcommand{\nn}{\nonumber \\}
\newcommand{\es}{& = &}
\newcommand{\rs}{\, = \,}
\newcommand{\ps}{& + &}
\newcommand{\ms}{& - &}
\newcommand{\ts}{& \times &}
\newcommand{\nt}{\nn \ts}
\newcommand{\np}{\nn \ps}
\newcommand{\nm}{\nn \ms}

\newcommand{\cA}{ {\cal A} }

\newcommand{\cD}{ {\cal D} }
\newcommand{\cM}{ {\cal M} }
\newcommand{\cH}{ {\cal H} }

\newcommand{\cG}{ {\cal G} }
\newcommand{\cT}{ {\cal T} }

\newcommand{\cU}{ {\cal U} }
\newcommand{\cL}{ {\cal L} }
\newcommand{\cP}{ {\cal P} }

\begin{document}
\title{    Fermion mass mixing and vacuum triviality in \\ 
           the renormalization group procedure for effective particles } 
\author{   Stanis{\l}aw D. G{\l}azek }
\affiliation{  Institute of Theoretical Physics,
               Faculty of Physics, 
               University of Warsaw }
\date{ 15 May, 2013 }
\begin{abstract}
Renormalization group procedure for effective particles is
applied to the model quantum theory of free fermions to
which one adds an interaction in the form of a mass mixing
term. If one used a standard approach based on the instant
form of dynamics, the theory would suffer from a generic
vacuum problem caused by a divergent production of virtual
quanta out of a bare vacuum and it would require an
adjustment of its degrees of freedom to the added
interaction term before quantization, considered a means of
avoiding the quantum vacuum problem. In the effective
particle approach, the quantum vacuum problem is dealt with
instead by using the front form of dynamics, where the pair
production is excluded by momentum conservation. The
corresponding Hamiltonian includes mass parameters through
constraint equations while the required quantum field
operators are constructed independently of all mass parameters, 
including the parameters that appear in the added mass
mixing interaction term. Then the masses and states of
physical fermions emerge at an end of the non-perturbative
calculation that is carried out entirely in one and the same
interacting quantum theory with a trivial vacuum and no
quantization adjustment. An a priori infinite set of
renormalization group equations for all momentum modes of
fermion fields is reduced to just one equation for a
two-by-two mass squared matrix, thanks to 7 kinematical
symmetries of the front form (the instant form has only 6).
For strong mass mixing interactions, the fermion model
solutions qualitatively differ from the analogous, earlier
found boson model solutions by the absence of tachyons.
\end{abstract} 
\maketitle
%
%
%
%
%
%
%
%

\section{ Introduction }
\label{I}

It has been recently shown~\cite{bosons} that a theory of
quantum scalar fields with mass mixing interactions can be
solved non-perturbatively using the renormalization group
procedure for effective particles (RGPEP). This article
shows that the RGPEP can also solve a quantum theory of
fermion fields with arbitrarily strong mass mixing
interactions. Thus, RGPEP is found to pass the test of
solving elementary theories beyond perturbative expansions.
For example, the procedure demonstrates that the fermion
theories do not have tachyon solutions no matter how strong
the mass-mixing interactions are, in distinction from the
boson theories that have tachyon solutions for sufficiently
strong mass-mixing interactions.

The theories of fermions or bosons with mass
mixing interactions can be constructed using
different forms of dynamics~\cite{DiracFF}. The
most commonly used form of dynamics is called the
instant form (IF), where the evolution of a system
is traced from one time to another.
Both the boson and the fermion theories exhibit an
ultraviolet divergent vacuum problem in their IF
versions. The problem is caused by a copious
creation of virtual particle pairs of unlimited
virtuality~\cite{DiracDeadWood}. The only IF
method that the author knows for circumventing the
vacuum problem caused by mass mixing is to return
to a classical theory and to change the
quantization procedure in a way that depends on
the mass-mixing interaction. This method is called
here {\it re-quantization}. Unfortunately, it is
not clear how to apply the method of
re-quantization to theories of great physical
interest, such as QCD or the electro-weak theory
with massive neutrinos. The reason is that the
corresponding relativistic quantum interactions
are not sufficiently understood to establish if
there exist some classical degrees of freedom that
are suitable for the purpose.

The RGPEP is systematically applied to a quantum theory with
mass mixing interactions without any need for
re-quantization. The vacuum problem is avoided in the RGPEP
by using an alternative form of dynamics to the IF. The
alternative form is called the front form (FF)~\cite{DiracFF}. 
In the FF, the evolution of a system is traced from one value 
of $t + z$ to another. The creation of pairs out of the bare
vacuum by a translation invariant interaction, which by
necessity conserves momentum, is not possible in a regulated
FF theory since the pairs must carry a non-zero kinematical
momentum while the vacuum carries zero. Moreover, the FF has
7 kinematical symmetries instead of only 6 in the IF and the
RGPEP preserves these 7 symmetries. The symmetries result in
a reduction of an infinite set of differential
renormalization group equations for all Fourier components
of quantum fields to just one $2 \times 2$ matrix equation
for masses of effective particles. This is not a mere
computational simplification because the reduction of
solving a mass-mixing theory to solving just two coupled
RGPEP equations for the masses of effective particles allows
one to avoid the issue of regulating the theory in
ultra-violet and subsequently removing the ultra-violet
cutoff dependence from observables, in order to recover the
Lorentz symmetry in the spectrum of solutions (see below).

Some details of construction of quantum fermion fields in
the RGPEP are worth mentioning because they are helpful in
handling the FF constraint equations and the Lorentz
symmetry. In distinction from the IF quantum fermion fields,
the unconstrained parts of the FF quantum fermion fields
have only two components instead of four and these two are
constructed purely kinematically in terms of their Fourier
components. Hence, the unconstrained quantum fields do not
depend on the fermion mass parameters. These parameters only
enter in the Hamiltonian as coefficients of products of the
quantum fields, as a consequence of the constraint
equations. Thus, the FF construction of quantum fermion
field operators avoids the IF difficulties due to assigning
masses to fermions as if they were free while the
Hamiltonian includes interactions. More generally, the
little group~\cite{Wigner} that preserves the front allows one to build
states and operators for fermions with arbitrary kinematical
momenta irrespective of the interaction. Such purely
kinematical construction of quantum field operators is not
possible in the IF theory. The reason is that the motion of
fermions in the IF requires a spinor representation of the
Lorentz boosts. A priori, the boost generators depend on
interactions and the simplest form of such dependence occurs
through the mass terms. This is also why the IF Fourier
expansion of quantum fermion fields depends on the fermion
masses. 

Comprehensive description of the RGPEP for fermions requires
several elements that are collected in several Appendices,
in order to avoid crowding the main text with details.
However, the main text does include the details that concern
basic features of quantum field theory, a subject of a long
history~\cite{QFT1,QFT2} and unyielding
relevance~\cite{DiracFoundations}. The only details that are
not discussed comprehensively concern regularization of
fermion fields. These details are not required for
completeness of the article because the RGPEP equations turn
out to reduce to two equations for mass parameters only and
these equations are entirely independent of particle momenta
in our mass mixing models (this is a consequence of a
general design of the RGPEP). Thus all regularizations based
on limiting a momentum range in the Fourier expansion of
quantum fields are of no consequence for the obtained
solutions and the solutions satisfy all requirements of
special relativity and quantum mechanics within any finite
range of momentum under consideration. 

So, the article point is not just that the RGPEP can be used
to solve a simple theory, but that it is defined in quite
general terms and works well in the test case with fermions,
in addition to the test it had passed earlier for
bosons~\cite{bosons}. Of course, solutions of the RGPEP
equations in complex theories cannot be found as easily as
in the simple models with mass mixing. Nevertheless, one may
hope that the RGPEP will help in searches for feasible ways
of constructing numerical approximations to solutions of
complex relativistic theories, such as the FF version of
QCD~\cite{Wilsonetal}.

The article is organized as follows. Section~\ref{IFtheory}
describes the model theory of fermions with mass mixing
interaction terms in the standard approach based on the IF
of dynamics. One starts with constructing a quantum theory
of free fermions, adds mass mixing interactions, discovers
the divergent vacuum problem, and goes back to the classical
theory in order to re-quantize it using new fields and thus
get rid of the vacuum problem. The outcome is an expectation
of how the solution to the quantum theory of fermions with
mass mixing interactions could look like. Section~\ref{FFm}
describes the FF approach. Once the FF quantum theory is
defined, by constructing the unconstrained quantum field
operators kinematically and taking into account the
constraint equations in constructing the FF Hamiltonian, the
vacuum problem is absent because the interaction terms do
not create pairs from the bare vacuum state. The RGPEP
procedure is then applied to solving the quantum theory
without any need for re-quantization. The procedure leaves
the trivial vacuum state unchanged. At the end of the RGPEP,
one arrives at the same spectrum of solutions as the one
expected on the basis of re-quantization in the IF.
Sec.~\ref{Conclusion} concludes the article by an
explanation of a qualitative difference between the fermion
and boson models concerning tachyons when the mass mixing
interactions are strong. Appendix~\ref{gamma} describes a
representation of $\gamma$ matrices that is useful in
constructing FF theories. Kinematical construction of the FF
quantum fermion fields is described in
Appendix~\ref{spinors}. For completeness,
Appendix~\ref{ARGPEP} recapitulates elements of the RGPEP in
general terms. Explicit solutions of the RGPEP equations in
the fermion model are described in
Appendix~\ref{Asolutions}.

\section{ IF theory of mass mixing }
\label{IFtheory}

We start our discussion with a brief recollection of the
commonly known IF theory of free fermion fields. For
simplicity, we explicitly consider just two fields. The
quantum theory is obtained by imposing anti-commutation
relations on the fields. Then we add the mass mixing
interaction term to the free Hamiltonian and thus obtain an
elementary example of the Dirac vacuum
problem~\cite{DiracDeadWood}. The problem is then dealt with
by going back to a classical theory and introducing two new
fermion fields for which the classical Lagrangian density
does not contain mass mixing terms. The FF approach based on
the RGPEP will be shown in Sec.~\ref{FFm} to be different. 

\subsection{ IF free fermions }
\label{IFfree}

Consider the Lagrangian density, 
\beq
\label{LIF}
\cL 
\es 
\bar \psi (i \partial \hspace{-5.5pt}/ - \mu) \psi 
+
\bar \phi (i \partial \hspace{-5.5pt}/ - \nu) \phi
\, ,
\eeq
for two types of fermion fields $\psi$ and $\phi$ 
with masses $\mu$ and $\nu$. Variation of the action 
$A = \int d^4x \, \cL$ with respect to $\bar \psi$ 
and $\bar \phi$ yields the Dirac equations of motion
\beq
\label{DE1}
(i \partial \hspace{-5.5pt}/  - \mu)\, \psi \es 0 \, , \\
(i \partial \hspace{-5.5pt}/  - \nu)\, \phi \es 0 \, .
\eeq
The corresponding IF Hamiltonian has the form
\beq
H \es \int d^3x \, \cT^{00} \, ,
\eeq
where $\cT^{00} = \cH$ denotes the energy density, 
i.e., the $\rho=\sigma=0$ component of the 
energy-momentum density tensor 
\beq
\cT^{\rho \sigma} 
\es
{\partial \cL \over \partial \partial_\rho
\psi_\alpha} \, \partial^\sigma \psi_\alpha
+
{\partial \cL \over \partial \partial_\rho
\phi_\alpha} \, \partial^\sigma \phi_\alpha
-g^{\rho \sigma} \, \cL \, .
\eeq
The resulting Hamiltonian reads
\beq
\label{HIF1}
H = \int d^3x \, 
\left[ \,
\psi^\dagger (i \vec \alpha \vec \partial + \beta \mu) \psi 
+
\phi^\dagger (i \vec \alpha \vec \partial + \beta
\nu) \phi \,
\right]
\, .
\eeq
For the purpose of constructing a quantum theory,
the Fourier decomposition of the fields at $x^0=0$ 
is arranged in the forms
\beq
\psi(\vec x \,) 
\es 
\sum_{~~~~ \mu p s} \hspace{-18pt}\int  \, 
\left[ u_{\mu p s} \, b_{\mu p s}         \, e^{ i\vec p \, \vec x} 
     + v_{\mu p s} \, d_{\mu p s}^\dagger \, e^{-i\vec p \, \vec x} \right] \, , \\
\phi(\vec x \,) 
\es 
\sum_{~~~~ \nu p s} \hspace{-18pt}\int  \, 
\left[ u_{\nu p s} \, b_{\nu p s}         \, e^{ i\vec p \, \vec x} 
     + v_{\nu p s} \, d_{\nu p s}^\dagger \, e^{-i\vec p \, \vec x} \right] \, .
\eeq
We explicitly explain the notation for the 
field $\psi$. Notation for $\phi$ is 
obtained by replacing the mass $\mu$ 
with $\nu$. 

The meaning of summing over spins and 
integrating over momentum is defined by 
\beq
\sum_{~~~~ \mu p s} \hspace{-18pt}\int 
\es
\sum_{s = \, \pm 1} \int { d^3 p \over
(2\pi)^3 2 E_{\mu p} } \, ,
\eeq
$E_{\mu p} = \sqrt{ \mu^2 + \vec p^{\,\, 2} }$, etc.
The subscript $\mu$ refers to the dependence on 
the mass parameter. The spinors are obtained by 
boosting spinors at rest (cf. Ref.~\cite{IF}, 
Chap. 3),
\beq
\label{umupsIF}
u_{\mu p s} \es B(\mu, \vec p\,) \, u_{\mu 0 s} \, , \\
\label{vmupsIF}
v_{\mu p s} \es B(\mu, \vec p\,) \, v_{\mu 0 s} \, , 
\eeq
where the boost matrix in the spinor representation, 
\beq
B(\mu, \vec p\,) \es {1  \over \sqrt{ 2 \mu 
(E_{\mu p} + \mu ) } } \, 
\left( p \hspace{-4.1pt}/ \beta + \mu \right) \, ,
\eeq
acts on the spinors that correspond to fermions 
at rest. In the representation of $\gamma$-matrices 
used in Ref.~\cite{IF}, see Eq.~(\ref{g0IF}) in
Appendix~\ref{gamma}, the spinors at rest are the 
ones given in Eqs.~(\ref{u0IF}) and (\ref{v0IF})
after multiplication by $\sqrt{ 2 \mu}$. 

The quantum field $\hat \psi$ is obtained from $\psi$ 
by replacing the Fourier coefficients $b$ and $d$ 
with operators. The non-zero anti-commutation 
relations the resulting operators satisfy read
\beq
\label{acrpsi}
\left\{ \hat \psi(\vec x\,), \hat \psi^\dagger(\vec x\, ') \right\}
\es 
\delta^3(\vec x - \vec x \, ') \, , \\
\label{acrbd}
\left\{ b_{\mu p s}, b^\dagger_{\mu p' s'} \right\}
\es
\left\{ d_{\mu p s}, d^\dagger_{\mu p' s'}
\right\} \nn
\es
2E_{\mu p} (2\pi)^3 \delta^3( \vec p - \vec p \, ') \, \delta_{s s'} \, .
\eeq
The quantum field $\hat \phi$ is obtained in 
a similar way keeping $\nu$ in place of $\mu$.

The quantum Hamiltonian takes the form 
\beq
\label{HIFQ1}
\hat H = \int d^3x \, 
{:} \left[ \,
\hat \psi^\dagger (i \vec \alpha \vec \partial + \beta \mu) \hat \psi 
+
\hat \phi^\dagger (i \vec \alpha \vec \partial + \beta
\nu) \hat \phi \,
\right]{:} \, , 
\eeq
where the symbols : denote normal ordering of the
operators between them, i.e., creation operators are 
put to the left of the annihilation operators. Such 
ordering involves dropping an infinite additive 
numerical constant of dimension energy from the 
Hamiltonian. 

To avoid the infinity, one would have to limit the
range of momentum in the Fourier expansion of the
quantum fields and the size of space volume in
which the theory is being constructed. On the
other hand, a numerical constant does not
contribute to the resulting quantum mechanics and
can be ignored. This is justified by saying (e.g.,
see Ref~\cite{Weinberg}, p. 297) that the
resulting quantum Hamiltonian has the structure 
\beq
\label{HIFQ2}
\hat H_0 \es 
\sum_{~~~~ \mu p s} \hspace{-18pt}\int 
\,
E_{\mu p} \,
\left( b^\dagger_{\mu p s}  b_{\mu p s}
     + d^\dagger_{\mu p s}  d_{\mu p s} \right)
\np
\sum_{~~~~ \nu p s} \hspace{-18pt}\int 
\,
E_{\nu p} \,
\left( b^\dagger_{\nu p s}  b_{\nu p s}
     + d^\dagger_{\nu p s}  d_{\nu p s} \right)
\, ,
\eeq
which is physically right for counting energy 
of free fermions. The subscript 0 is used to 
indicate that there is no interaction.

All the relations given above are commonly known. They are
given here for the purpose of observing that the
construction of quantum fields in the IF of dynamics relies
on the representation of boosts for fermions that is valid
only if they are free. The issue is that in a theory with
interactions the complete boost operators depend on the
interactions. The boosts do not belong to the little
group~\cite{Wigner} associated with a time-like four-vector
$n$ that defines the canonical quantization hyperplane in
space-time through condition $nx=x^0=0$, where $x$ denotes
the co-ordinates of points in space-time in the frame of
reference of an observer who carries out the quantization
procedure and whose world-line lies along $n$. The general
feature of boosts depending on interaction is also exhibited 
in the case of the mass mixing interaction to be discussed 
below. Not only the mass parameters must be chosen properly 
in the IF quantization of fields but also the quantum creation 
and annihilation operators need proper definitions. Such 
definitions are necessary in order to avoid the IF Dirac 
vacuum problem~\cite{DiracDeadWood} described in Sec.~\ref{vacuum}
below. In general, however, one does not know what mass
parameters and operators to assign to fermions in the IF
construction of a quantum field theory in the presence of
interactions, especially in the case of strong interactions
to which one cannot apply any perturbative procedure that
starts from the free particle approximation. The ultimate
difficulty with the free fermion mass assignment is
encountered in the case of confined quarks. It is hence
helpful to keep in mind while following further discussion
of the theory of fermions with mass mixing in the IF of
dynamics that the FF construction of the theory is different
and does not require any assignment of masses to fermions in
the definition of quantum field operators on the front where
the initial conditions are specified. 

\subsection{ IF mass mixing and the vacuum }
\label{vacuum}

The Lagrangian density including the 
mass mixing interaction is defined by
writing
\beq
\label{LIFI}
\cL 
\es 
\bar \psi (i \partial \hspace{-5.5pt}/ - \mu) \psi 
+
\bar \phi (i \partial \hspace{-5.5pt}/ - \nu) \phi
- m \left( \bar \psi \phi + \bar \phi \psi \right)
\, .
\eeq
The corresponding quantum Hamiltonian 
of canonical IF quantization procedure 
(e.g., see Refs.~\cite{IF,Weinberg,Peskin}) 
is
\beq
\hat H \es \hat H_0 + \hat H_I \, ,
\eeq
where $\hat H_0$ is given in Eq.~(\ref{HIFQ2}) 
and the interaction term reads
\beq
\hat H_I \es m \int d^3x \, : \left( \hat \psi^\dagger \gamma^0  \hat \phi 
         + \hat \phi^\dagger \gamma^0  \hat \psi \right) : \, .
\eeq
Using the Fourier expansions for the 
quantum fields described in previous 
section, integrating over space and 
performing normal ordering, one obtains
\beq
\label{HIIF}
\hat H_I 
\es
m 
\sum_{~~~~ \mu p  s} \hspace{-18pt}\int \,\,\,  
\sum_{s'} {1 \over 2 E_{\nu p} }
\left[
\bar u_{\mu p  s } \, u_{\nu p  s'} \, 
b_{\mu p  s }^\dagger \, b_{\nu p  s'} 
\right.
\np
\left.
\bar u_{\mu p  s } \, v_{\nu -p s'} \, 
b_{\mu p  s }^\dagger \, d_{\nu -p s'}^\dagger
+
\bar v_{\mu p  s } \, u_{\nu -p s'} \, 
d_{\mu p  s }         \, b_{\nu -p s'}  
\right.
\nm
\left.
\bar v_{\mu p  s } \, v_{\nu p s'} \, 
d_{\nu p s'}^\dagger \, d_{\mu p  s } 
\right]
+
(\mu \leftrightarrow \nu) \, .
\eeq
The Hamiltonian $\hat H$ can be considered an
operator in the Fock space whose basis states 
are created from the bare vacuum state $|0\rangle$
by products of creation operators. The state 
$|0\rangle$ is defined by the condition that it 
is annihilated by all annihilation operators
in the theory.

Unfortunately, the interaction Hamiltonian $\hat H_I$
is able to copiously create fermion-anti-fermion
pairs from the bare vacuum state $|0\rangle$ no
matter how small the mass mixing parameter $m$ is.
Such creation leads to the divergences that were
considered severe enough to question the existence
of the Schr\"odinger picture in QED~\cite{DiracDeadWood}. 
Indeed, the vacuum problem in the mass mixing model 
is an elementary example of the general vacuum problem 
in relativistic quantum field theory with interactions. 
The general vacuum problem has a long history of 
attempts to solve it, motivated by its basic 
significance in physics. The literature concerning 
the problem is very rich. The list of Refs.~\cite{Nambu, 
DiracDeadWood, GOR, tHooftVeltman, KogutSusskind, 
CasherSusskind, confinement, SVZ, FeynmanQCD2, 
SSbreaking, GasserLeutwyler, condensates, 
WeinbergCosmology1, Wilsonetal, Marisqq, 
BrodskyPauliPinsky, npRGPEP, WeinbergCosmology2, 
BrodskyRobertsSchrockTandy} amply illustrates this 
statement. Despite that the list is greatly incomplete 
and partly biased by the stress on works that concern
differences between formulations of the vacuum problem 
in the IF and FF of dynamics, the quoted works are 
indicative of the development of ideas concerning 
the vacuum problem over recent half of a century.

Our further discussion is limited to the simple
mass mixing model. We proceed to an explanation 
of the divergences that appear in its vacuum 
problem.

Consider the pair-creation term in $\hat H_I$
of Eq.~(\ref{HIIF}),
\beq
\label{hIF}
\hat h
\es 
m 
\sum_{~~~~ \mu p  s} \hspace{-18pt}\int \,\,\,  \sum_{s'} {1 \over 2 E_{\nu p} }
\,
\bar u_{\mu p  s } \, v_{\nu -p s'} \, 
b_{\mu p  s }^\dagger \, d_{\nu -p s'}^\dagger \, .
\eeq
This term is analogous to the model Hamiltonian 
term of Eq.~(9) in Ref.~\cite{DiracDeadWood}. 
The term $\hat h$ contains the spinor product
\beq
\bar u_{\mu p  s } \, v_{\nu -p s'} 
\es
\left(
\sqrt{ E_{\nu p} + \nu \over  E_{\mu p} + \mu }
+
\sqrt{ E_{\mu p} + \mu \over  E_{\nu p} + \nu }
\, \right)
\nt
\chi_s^\dagger \, \vec \sigma \vec p \, i \sigma^2
\, \chi_{s'} \, ,
\eeq
where the two-component spinors $\chi_s$ are the
ones introduced in spinors of fermions at rest, 
$u_{\mu 0 s}$ and $v_{\mu 0 s}$ in Eqs.~(\ref{umupsIF})
and (\ref{vmupsIF}). 

The eigenvalue problem for the ground state 
of $\hat H$ involves $\hat h$. The state 
obtained by acting with $\hat h$ on the bare 
vacuum, 
\beq
|h\rangle \es \hat h |0\rangle \, ,
\eeq
differs from the bare vacuum. The question 
arises how to find the true ground state of 
the theory, if it is not $|0\rangle$.

If a part $\hat h$ of $\hat H_I$ produces $|h\rangle 
\neq 0$, the ground eigenstate of $\hat H$ must 
involve the component proportional to $|h\rangle$ 
once it contains a component proportional to 
$|0\rangle$. Then the term $\hat h^\dagger$ in 
the same $\hat H_I$ produces a state of an infinite 
norm when acting on $|h\rangle$. Further action of 
$\hat h$ and $\hat h^\dagger$ produces states with 
additional pairs and infinities. The ground state 
would have to involve some combination of all of 
them. Ref.~\cite{DiracDeadWood} points out that the
problem with ultra-violet divergences in all these 
states leads to violation of the Lorentz symmetry in 
a mathematically well-defined theory. In our example, 
the corresponding reasoning could go as follows.

Since $|h\rangle$ is an eigenstate of the three-momentum 
operator with eigenvalue 0, it has, as all eigenstates 
of the three-momentum operator, a norm squared proportional 
to the volume of space, or $V = \int d^3 x = (2\pi)^3 
\delta^3(0)$. This is a general feature and it does not 
pose serious problems for application of a theory to the 
description of physical phenomena of a finite size. 
However, a direct evaluation yields
\beq
\langle h | h \rangle 
\es
\langle 0 | \hat h^\dagger \hat h |0 \rangle \\
\es
V m^2 
\sum_{~~~~ \mu p  s} \hspace{-18pt}\int \,\,\,  
\sum_{s'} {1 \over 2 E_{\nu p} }
| \bar u_{\mu p  s } \, v_{\nu -p s'} |^2 \\
\es
V m^2 
\int { d^3 p \over (2\pi)^3 2 E_{\mu p} } \, 
{\vec p\,^2 \over E_{\nu p} } \, .
\eeq
This result means that the norm of $|h\rangle$ is 
infinite unless the number of momentum states of 
a single fermion in the theory is limited by some 
ultraviolet cutoff, say $\Lambda$, on $|\vec p\,|$.
Otherwise, action of $\hat H_I$ takes states out 
of the Hilbert space. To obtain a mathematically 
acceptable theory, the Fourier expansion of the 
fermion fields $\hat \psi(\vec x\,)$ and $\hat 
\phi(\vec x\,)$ at $x^0=0$ must be cut off at some 
finite $\Lambda$, or regulated in some other way 
in the ultra-violet so that the range of momenta
is effectively limited by some $\Lambda$. But 
every finite cutoff $\Lambda$ on particle momenta 
violates the Lorentz symmetry. Since this symmetry 
is believed to be physically valid to a great 
precision, the theory with a finite cutoff faces
the problem of applicability in physics. In 
particular, the theoretical assumption that there 
exists a vacuum state that is invariant with 
respect to the Lorentz transformations is not
compatible with a precisely defined theory.

The ultra-violet divergent pair creation that
causes the vacuum problem also leads to
divergences in other states and the Schr\"odinger
evolution operator $\exp(-i \hat H t)$ cannot be
understood as an operator in terms of the 
corresponding Taylor series acting on any state. 
The question then arises if a relativistic quantum
theory with a mass mixing interaction can be
formulated in the IF of quantum Hamiltonian
dynamics. The positive answer to this question
that is discussed below in
Sec.~\ref{re-quantization} involves a well-known
procedure that we call the IF re-quantization.
However, although the re-quantization works for
the mass mixing model, it does not tell us at all
how to seek a solution of the general vacuum problem 
in other theories, as the half of a century of
research we refer to above attests. The
alternative approach that is based on the RGPEP,
and can be employed to study also other theories,
will be discussed later on in Sec.~\ref{FFm}.

\subsection{ IF re-quantization }
\label{re-quantization}

The Lagrangian density of Eq.~(\ref{LIFI}) 
can be written in the equivalent form
\beq
\label{cLPsi}
\cL  \es \bar \Psi ( i \partial \hspace{-5.5pt}/  - M) \Psi \, , 
\eeq
where the field $\Psi$ is a double size 
fermion field built from the two four-component 
fields $\psi$ and $\phi$,
\beq
\Psi \es \left[ \begin{array}{c} \psi \\ 
\phi \end{array} \right] \, , 
\eeq
so that $\Psi$ has altogether 8 components.
The mass symbol $M$ stands for the $8 \times 8$ 
mass matrix, formed out of four $4 \times 4$ 
unit matrices multiplied each by $\mu$, $\nu$, 
or $m$,
\beq
\label{M}
M    \es \left[ \begin{array}{cc} \mu & m \\ 
                m & \nu \end{array} \right] \, .
\eeq
Let the notation be arranged so that $\mu - \nu > 
0$. This is always possible except for the case 
of fermions with equal masses, i.e., $\mu = \nu$, 
which is special and will be commented on separately
in further discussion. The eigenvalues and normalized 
eigenvectors of the mass matrix $M$ are
\beq
\label{m12}
m_{1,2}
\es \left[ \, \mu + \nu  \, \pm \, 
             (\mu - \nu) \, \epsilon \, \right]/2 \, , \\
\label{v12}
v_1 \es \left[ \begin{array}{r} \cos \varphi  \\
                              - \sin \varphi
               \end{array} \right] \, , 
\quad \quad 
v_2 \rs \left[ \begin{array}{r} \sin \varphi  \\
                                \cos \varphi
               \end{array} \right] \, , 
\eeq 
where 
\beq
\label{epsilon}
\epsilon \es   \sqrt{1 + [2 m/(\mu-\nu)]^2} \, , \\
\label{angle}
\varphi  \es - \arctan{ \sqrt{ \epsilon - 1 \over 
                               \epsilon + 1} } \, .
\eeq
The double size fermion field $\Psi$ can be 
written in terms of two new four-component fields 
$\psi_1$ and $\psi_2$ using the eigenvectors of $M$,
\beq
\label{PsiDublet}
\Psi
\es
\psi_1 \, v_1 + \psi_2 \, v_2 \, .
\eeq
The new four-component fermion fields are  
\beq
\label{xi1}
\psi_1 \es \cos \varphi \,\, \psi -  \sin \varphi \,\, \phi \, , \\
\label{zeta1}
\psi_2 \es \sin \varphi \,\, \psi +  \cos \varphi \,\, \phi  \, .
\eeq
The IF re-quantization is based on expressing the
classical Lagrangian density of Eq.~(\ref{LIFI}) 
in terms of the fields $\psi_1$ and $\psi_2$. Since 
these fields multiply the orthogonal eigenvectors 
of $M$, they are multiplied in the Lagrangian by the 
corresponding eigenvalues $m_1$ and $m_2$ and they 
are not mixed by $M$. Since the Lagrangian density 
term with $i \partial \hspace{-5.5pt}/$ is the same 
for both fields $\psi$ and $\phi$ and does not mix 
them, the orthogonal rotation of fields from $\psi$ 
and $\phi$ to $\psi_1$ and $\psi_2$ does not alter 
this term.

The Lagrangian density of Eq.~(\ref{LIFI}) takes
the form 
\beq
\label{Lreq}
\cL 
\es 
\bar \psi_1 (i \partial \hspace{-5.5pt}/ - m_1) \psi_1 
+
\bar \psi_2 (i \partial \hspace{-5.5pt}/ - m_2) \psi_2 \, .
\eeq
One can now quantize the independent fields $\psi_1$ 
and $\psi_2$ as if they were free, because there is 
no interaction between them; the mass mixing is 
removed at the classical level of dealing with the 
fields. The only effect of the original mass mixing 
interaction is that the masses $m_1$ and $m_2$ are 
the eigenvalues of $M$. We call this new quantization 
a re-quantization because it removes the mass mixing 
interaction terms that caused trouble in the original 
quantum theory of fields $\hat \psi$ and $\hat \phi$. 
We have stepped back to the classical theory, introduced
new field variables $\psi_1$ and $\psi_2$, and now we
construct the new quantum operators $\hat \psi_1$ and 
$\hat \psi_2$ instead of struggling with the old ones 
$\hat \psi$ and $\hat \phi$.

The quantum operators $\hat \psi_1$ and $\hat \psi_2$ 
are obtained by imposing standard anti-commutation 
relations of the type indicated in Eqs.~(\ref{acrpsi}) 
and (\ref{acrbd}). Following the same steps that
previously led to Eq.~(\ref{HIFQ2}), one now obtains
\beq
\label{reqHIFQ2}
\hat H \es 
\sum_{~~~~~~ m_1 p s} \hspace{-25pt} \int 
\,
E_{m_1 p} \,
\left( b^\dagger_{m_1 p s}  b_{m_1 p s}
     + d^\dagger_{m_1 p s}  d_{m_1 p s} \right)
\np
\sum_{~~~~~~ m_2 p s} \hspace{-25pt} \int 
\,
E_{m_2 p} \,
\left( b^\dagger_{m_2 p s}  b_{m_2 p s}
     + d^\dagger_{m_2 p s}  d_{m_2 p s} \right)
\, ,
\eeq
which is a quantum IF Hamiltonian for two types
of free fermions with masses $m_1$ and $m_2$.
The vacuum problem appears now absent because 
the mixing is classically included in the new
mass parameters and the re-quantized theory 
does not produce terms of the type $b^\dagger 
d^\dagger$ and $d \, b$ any more.

The situation is similar to the one in scalar 
theory with mass mixing interactions discussed 
in Ref.~\cite{bosons}. Disappearance of terms 
such as $b^\dagger d^\dagger$ results from the 
choice of masses in $E_{m_1 p}$ and $E_{m_2 p}$. 
However, instead of using these energies for 
constructing the time derivatives of fields that 
play the role of canonical momenta, one constructs 
the corresponding spinors whose matrix elements 
in front of the terms such as $b^\dagger d^\dagger$
vanish. 

As in the scalar case, the author does not know 
of any practical extension of the IF re-quantization 
recipe for fermion mass mixing that could be 
systematically applied in relativistic theories 
with other interactions beyond the perturbative 
expansion that is based on a free particle 
approximation with nearly precise match between 
the theoretical Lagrangian mass parameters and 
masses of physical particles. The RGPEP will be 
shown below to deal with the mass mixing 
interaction quite differently, entirely
within a quantum theory of $\hat \psi$ and $\hat
\phi$, i.e., without a need to define new fields
$\psi_1$ and $\psi_2$ and quantizing them from
scratch to define $\hat \psi_1$ and $\hat \psi_2$. 
This means that the RGPEP works in a way that can 
be systematically tried also in application to 
other types of interaction than just the mass 
mixing.

\section{ FF theory of mass mixing }
\label{FFm}

The FF of dynamics aims at description of the 
evolution of a system from one hyperplane of
constant $x^+ = x^0 + x^3$ to 
another~\cite{DiracFF}, with the front $x^+ 
= 0$ used to set up a quantum theory. We use
notation $v^\pm = v^0 \pm v^3$ and $v^\perp
= (v^1, v^2)$ for all four-vectors. The same
convention is adopted for denoting components 
of all tensors. In particular, $\partial^\pm
= 2 \,\partial/\partial x^\mp$ and $\partial^\perp
= - \partial/\partial x^\perp$.

In the FF of dynamics, it is useful to consider 
the Lagrangian density of Eq.~(\ref{LIFI}) in
the form of Eq.~(\ref{cLPsi}). The Euler-Lagrange 
equations read
\beq
( i \partial \hspace{-5.5pt}/  - M) \, \Psi \es 0 \, .
\eeq
Using conventions described in Appendix \ref{gamma}, 
one can write these equations as 
\beq
\label{projectedM}
i \partial^- \Psi_+ + i \partial^+ \Psi_- 
- 
( i \alpha^\perp \partial^\perp + \beta M) (\Psi_+ + \Psi_-) 
\es 0 \, .
\nn
\eeq
Projection with $\Lambda_+$ yields equations of 
motion that involve $\partial^-$,
\beq
\label{projectedM2}
i \partial^- \Psi_+ 
\es 
( i \alpha^\perp \partial^\perp + \beta M) \Psi_-  \, .
\eeq
Projection with $\Lambda_-$ produces complementary 
constraints, i.e., equations that do not involve 
$\partial^-$,
\beq
\label{projectedM1}
i \partial^+ \Psi_- 
\es 
( i \alpha^\perp \partial^\perp + \beta M) \Psi_+ \, .
\eeq

In deriving the corresponding FF Hamiltonian, 
one can take advantage of Refs.~\cite{Yan1,Yan2}
and obtain
\beq
P^- \es {1 \over 2 }\int dx^- d^2x^\perp \, \cT^{+-} \, ,
\eeq 
with the energy-momentum density component
\beq
{ 1\over 2} \cT^{+-} 
\es \Psi_+^\dagger i \partial^- \Psi_+ \\
\es
\Psi_+^\dagger 
( i \alpha^\perp \partial^\perp + \beta M)  
{ 1 \over i \partial^+} \,  
( i \alpha^\perp \partial^\perp + \beta M) \Psi_+
\, .
\nn
\eeq
The density involves the non-local inverse 
of the differential operator,
\beq
{ 1 \over i \partial^+ } f(x^-, x^\perp)
\es
{1 \over 2} \left( \int_{-\infty}^{x^-} 
                 - \int_{x^-}^{+\infty} \right) 
            dy^- \, f(y^- , x^\perp) \, .
\nn
\eeq
For finite and non-zero momentum arguments 
of the Fourier transform $\hat f(p^+, p^\perp)$ 
of the function $f(x^-,x^\perp)$ that vanishes 
at the FF ``spatial'' infinity, this operation 
means simply dividing by $p^+$. It will be shown 
below that the RGPEP equations in the mass mixing 
model are completely independent of the momentum 
variables $p^+$ and $p^\perp$. Therefore, one 
does not have to deal here with subtle aspects 
of modes with $p^+=0$. 

Having accepted the inverse of $i \partial^+$ 
as a division of the Fourier components by 
their $p^+$, one has
\beq
\label{Pminus}
P^- \es \int dx^- d^2x^\perp \, 
\Psi_+^\dagger \, { - \partial^{\perp \, 2} + M^2  
\over i \partial^+} \,  \Psi_+  \, ,
\eeq 
where
\beq
\label{Msquared}
M^2    
\es 
\left[ \begin{array}{cc} \mu^2 + m^2  & m(\mu+\nu) \\ 
                         m(\mu + \nu) & \nu^2 + m^2 \end{array} \right]
\, .
\eeq
The next step is to define the 
corresponding quantum theory.

\subsection{ FF quantization }
\label{FFquantization}

The quantum Hamiltonian $\hat P^-$ defined by 
\beq
\label{P-1}
\hat P^- \es \int dx^- d^2x^\perp \, 
: \hat \Psi_+^\dagger \, { - \partial^{\perp \, 2} + M^2  
\over i \partial^+} \,  \hat \Psi_+ : \, ,
\eeq 
can be obtained by using the representation
of the $\gamma$ matrices defined in 
Appendix~\ref{gamma} and taking advantage
of the results for spinors and quantization 
of a fermion field in Appendix~\ref{spinors}.

In analogy to Eq.~(\ref{psiqf}), one can 
write classical fields at $x^+=0$ in 
the form
\beq
\label{psiqftext}
\psi(x) \es \left[ \begin{array}{c} \zeta(x) \\ 
                                      \xi(x)
\end{array}\right] \, , \\
\label{phiqftext}
\phi(x) \es \left[ \begin{array}{c} \omega(x) \\ 
                                      \rho(x) \end{array}\right] \, .
\eeq
In the representation of $\gamma$-matrices 
defined in Eqs.~(\ref{g0FF}) and (\ref{g12FF}), 
one has
\beq
\label{psi+qftext}
\psi_+(x) \es \left[ \begin{array}{c} \zeta(x) \\
0 \end{array}\right] \, ,
\quad 
\phi_+(x) \rs \left[ \begin{array}{c} \omega(x) \\
 0 \end{array}\right] \, ,
\eeq
so that the double size fermion field $\Psi_+$  
is composed of the two-component fermion fields 
$\zeta(x)$ and $\omega(x)$ according to
\beq
\label{Psi+text}
\Psi_+(x) \es \left[ \begin{array}{c} \zeta(x) \\
                                      0        \\ 
                                     \omega(x) \\
                                      0 \end{array}\right] \, .
\eeq
The quantum fields $\hat \psi_+$ and $\hat \phi_+$ 
are obtained by changing the classical fields 
$\zeta(x)$ and $\omega(x)$ to operators according
to the pattern of Eq.~(\ref{zetaqf}), with
\beq
\label{zetaqftext}
\hat \zeta(x) 
\es 
\sum_{~~ps} \hspace{-13pt}\int  \, \sqrt{p^+} \,
\left[  b_{\zeta ps}  \, e^{-ipx} 
      - d_{\zeta ps}^\dagger \, e^{ ipx} \sigma^1 \right] \, \chi_s  \, , \\
\label{omegaqftext}
\hat \omega(x) 
\es 
\sum_{~~ps} \hspace{-13pt}\int  \, \sqrt{p^+} \,
\left[  b_{\omega ps}  \, e^{-ipx} 
      - d_{\omega ps}^\dagger \, e^{ ipx} \sigma^1 \right] \, \chi_s  \, , 
\eeq
where 
\beq
\label{notationqft}
\sum_{~~ps} \hspace{-13pt}\int 
\es
\sum_{s = \, \pm 1} \int_{-\infty}^{+\infty}  { d^{2} p^\perp \over (2\pi)^2} \,
\int_0^{+\infty}  {d p^+ \over 2(2\pi) p^+} 
\eeq
and the operators $b_{\zeta ps}$, $d_{\zeta ps}$,
$b_{\omega ps}$, and $d_{\omega ps}$, annihilate 
fermions and anti-fermions of two kinds, 
respectively. The non-zero canonical 
anti-commutation relations at $x^+=0$, 
\beq
\label{crqftext}
\left\{ \hat \zeta(x),  \hat \zeta^\dagger(x') \right\}
\es
\left\{ \hat \omega(x), \hat \omega^\dagger(x') \right\}
\rs 
\delta^3(x - x') \, , 
\eeq
correspond to  
\beq
\label{bdzeta}
\left\{ b_{\zeta ps}, b^\dagger_{\zeta p's'} \right\}
\es
\left\{ d_{\zeta ps}, d^\dagger_{\zeta p's'} \right\}
\nn
\es
2p^+ (2\pi)^3 \delta^3(p - p') \, \delta_{s s'} \, , \\
\label{bdomega}
\left\{ b_{\omega ps}, b^\dagger_{\omega p's'} \right\}
\es
\left\{ d_{\omega ps}, d^\dagger_{\omega p's'} \right\}
\nn
\es
2p^+ (2\pi)^3 \delta^3(p - p') \, \delta_{s s'} \, .
\eeq
The above operator representations of 
quantum fermion fields $\hat \zeta(x)$
and $\hat \omega(x)$ at $x^+=0$ are 
universal in the sense that they are 
independent of the fermion mass parameters.

In terms of the quantum fields $\hat \zeta(x)$
and $\hat \omega(x)$, the Hamiltonian of 
Eq.~(\ref{P-1}) reads
\beq
\label{P-2}
\hat P^- 
\es 
\hat P_f^- + \hat P_I^- \, , 
\eeq
where the free Hamiltonian is
\beq
\hat P_f^-
\es
\int dx^- d^2x^\perp \, 
: \left( 
\hat \zeta^\dagger \, { - \partial^{\perp \, 2} + \mu^2  
\over i \partial^+} \,  \hat \zeta 
\right.
\np
\left.
\hat \omega^\dagger \, { - \partial^{\perp \, 2} + \nu^2  
\over i \partial^+} \,  \hat \omega 
\right) : \, , 
\eeq
and the interaction Hamiltonian is
\beq
\hat P_I^-
\es
\int dx^- d^2x^\perp \, 
: \left[
\hat \zeta^\dagger \, { m(\mu + \nu) \over i \partial^+} \,  \hat \omega 
\right.
\np
\left.
\hat \omega^\dagger \, { m(\mu + \nu) \over i \partial^+} \,  \hat \zeta 
+
\hat \zeta^\dagger \, { m^2 \over i \partial^+} \, \hat \zeta 
+
\hat \omega^\dagger \, { m^2 \over i \partial^+} \, \hat \omega 
\right] :  \, .
\nn
\eeq 
The interaction contains terms linear in 
the mass mixing parameter $m$ and terms 
quadratic in $m$. The latter appear because 
of the FF constraint equations. 

Evaluation of the Hamiltonian in terms of 
the creation and annihilation operators yields
\beq
\label{P-bd}
\hat P^-
\es 
\sum_{~~ps} \hspace{-12pt}\int \,
  \left[ \left( p_\mu^- + { m^2 \over p^+} \right) 
  \,\left( b^\dagger_{\zeta p s} \, b_{\zeta p s} 
         + d^\dagger_{\zeta p s} \, d_{\zeta p s} \right)
\right.
\np
\left.
         \left( p_\nu^- + { m^2 \over p^+} \right) 
  \,\left( b^\dagger_{\omega p s} \, b_{\omega p s} 
         + d^\dagger_{\omega p s} \, d_{\omega p s} \right)
\right.
\np
\left.
        { m(\mu + \nu) \over p^+} \,
  \, \left( b_{\zeta  p s}^\dagger \, b_{\omega p s}   
          + d_{\omega p s}^\dagger \, d_{\zeta  p s} 
\right.
\right.
\np
\left.
\left.
            b_{\omega p s}^\dagger \, b_{\zeta  p s}   
          + d_{\zeta  p s}^\dagger \, d_{\omega p s} \right)
\right] \, ,
\eeq
where 
\beq
p_\mu^- \es { p^{\perp \, 2} + \mu^2 \over p^+ }
\, , \quad 
p_\nu^- \rs { p^{\perp \, 2} + \nu^2 \over p^+ } 
\, .
\eeq
Note that the FF condition that all quanta have
positive momentum $p^+$ eliminates terms of the
type $b^\dagger d^\dagger$ and $b d$. The negative
sign in front of $d^\dagger$ in $\hat \zeta$ 
of Eq.~(\ref{zetaqftext}) and $\hat \omega$ of
Eq.~(\ref{omegaqftext}) is compensated by the sign 
of inverse of $i\partial^+$ and the normal ordering
of anti-fermion operators compensates the negative
signs in front of $d$ in $\hat \zeta^\dagger$ and 
$\hat \omega^\dagger$. The vacuum problem is thus
eliminated from the quantum theory. However, the
mass mixing interaction is still present in the 
FF Hamiltonian of Eq.~(\ref{P-bd}). This Hamiltonian 
provides the initial condition for the RGPEP. 

\subsection{ Application of the RGPEP }
\label{application}

The Hamiltonian $\hat P^-$ of Eq.~(\ref{P-bd})
is now considered an initial condition,
\beq
\cP^-_0 \es \hat P^- \, , 
\eeq
in the RGPEP scale evolution of $\cP_t^-$ 
according to the equation (see Appendix~\ref{ARGPEP})
\beq 
\label{tnpRGPEP}
{\cP^-_t}' 
\es
\left[ [ \cP^-_f, \cP^-_{Pt} ], \cP^-_t \right] \, .
\eeq 
The prime denotes differentiation with respect
to the scale parameter $t$ that ranges from 
0 at the beginning and tends to $\infty$ at the 
end of the RGPEP evolution. The equation is further 
explained in Appendix~\ref{ARGPEP}. Equation~(\ref{tnpRGPEP}) 
is the same general RGPEP equation that is used in 
the case of boson mass mixing and can also be used 
in other quantum field theories. 

Direct inspection of how Eq.~(\ref{tnpRGPEP}) works 
in the fermion mass mixing model (see below) allows 
one to write a general solution for $\cP_t^-$ in the 
form
\beq
\label{Pt}
\cP^-_t 
\es
\sum_{~~ps} \hspace{-12pt}\int \,
  \left[ A_{tp} 
  \, \left( b^\dagger_{\zeta p s} \, b_{\zeta p s} 
          + d^\dagger_{\zeta p s} \, d_{\zeta p s} \right)
\right.
\np
\left.
         B_{tp}  
  \, \left( b^\dagger_{\omega p s} \, b_{\omega p s} 
          + d^\dagger_{\omega p s} \, d_{\omega p s} \right) 
\right.
\np
\left.
        C_{tp} 
  \, \left( b_{\zeta  p s}^\dagger \, b_{\omega p s}    
          + b_{\omega p s}^\dagger \, b_{\zeta  p s}
\right.
\right.
\np
\left.
\left.
            d_{\zeta  p s}^\dagger \, d_{\omega p s}  
          + d_{\omega p s}^\dagger \, d_{\zeta  p s} \right)
\right] \, , 
\eeq
where the spin-independent coefficients are
\beq
\label{At}
A_{tp}    \es { p^{\perp \, 2} + \mu_t^2 \over p^+ } \, , \\ 
\label{Bt}
B_{tp}    \es { p^{\perp \, 2} + \nu_t^2 \over p^+ } \, , \\
\label{Ct}
C_{tp}    \es {                    m_t^2 \over p^+ } \, ,
\eeq
and the initial conditions at $t=0$ read
\beq
\label{mu0}
\mu_0^2 \es \mu^2 + m^2 \, , \\ 
\label{nu0}
\nu_0^2 \es \nu^2 + m^2 \, , \\
\label{m0}
m_0^2   \es m(\mu+\nu) \, .
\eeq
Note that the negative initial mixing term 
coefficient $m$ implies a negative initial 
value of $m_t^2$, which means that the 
notation $m_t^2$ is merely a formal 
indication that its dimension is mass 
squared but the value can be negative. 

\subsubsection{ Boost invariance }
\label{boostinvariance}

We explain how the design of the RGPEP leads to
boost invariant evolution equations for mass
parameters alone, which happens because the RGPEP
preserves all kinematical symmetries of the FF and
the mass mixing interactions in $\cP^-_t$ are
sufficiently simple. The formal features described
below are shown in Sec.~\ref{spectrum} to lead to
the Lorentz symmetry in the spectrum of solutions
in the model fermion theory.
 
According to the general RGPEP rules described
in Appendix~\ref{ARGPEP}, the operators $\cP^-_f$ 
and $\cP^-_{Pt}$ in Eq.~(\ref{tnpRGPEP}) are
\beq
\cP_f^-
\es 
\sum_{~~ps} \hspace{-12pt}\int \,
  \left[   p^-_\mu
  \,\left( b^\dagger_{\zeta p s} \, b_{\zeta p s} 
         + d^\dagger_{\zeta p s} \, d_{\zeta p s} \right)
\right.
\np
\left.
           p^-_\nu
  \,\left( b^\dagger_{\omega p s} \, b_{\omega p s} 
         + d^\dagger_{\omega p s} \, d_{\omega p
s} \right) \right] \, , \\
\cP^-_{Pt} 
\es
\sum_{~~ps} \hspace{-12pt}\int \, p^{+ \, 2}\,
  \left[ A_{tp} 
  \, \left( b^\dagger_{\zeta p s} \, b_{\zeta p s} 
          + d^\dagger_{\zeta p s} \, d_{\zeta p s} \right)
\right.
\np
\left.
         B_{tp}  
  \, \left( b^\dagger_{\omega p s} \, b_{\omega p s} 
          + d^\dagger_{\omega p s} \, d_{\omega p s} \right) 
\right.
\np
\left.
        C_{tp}  
     \left( b_{\zeta  p s}^\dagger \, b_{\omega p s}    
          + b_{\omega p s}^\dagger \, b_{\zeta  p s}
\right.
\right.
\np
\left.
\left.
            d_{\zeta  p s}^\dagger \, d_{\omega p s}  
          + d_{\omega p s}^\dagger \, d_{\zeta  p s} \right)
 \right] \, ,
\eeq
The resulting RGPEP generator is 
\beq
\label{generator}
{[} \cP_f^-, \cP_{Pt}^- {]}
\es
\sum_{~~ps} \hspace{-12pt}\int \, 
        C_{tp} \, p^{+ \, 2}(p_\mu^- - p_\nu^- ) \, 
 \left( b_{\zeta  p s}^\dagger \, b_{\omega p s}
\right.
\nm
\left.    
        b_{\omega p s}^\dagger \, b_{\zeta  p s}
      + d_{\zeta  p s}^\dagger \, d_{\omega p s}  
      - d_{\omega p s}^\dagger \, d_{\zeta  p s} \right)
\, .
\eeq
Consequently, Eq.~(\ref{tnpRGPEP}) reads
\beq
\label{rgpep}
{\cP^-_t}' 
\es
\sum_{~~ps} \hspace{-12pt}\int \,
  \left[ A_{tp}' 
  \, \left( b^\dagger_{\zeta p s} \, b_{\zeta p s} 
          + d^\dagger_{\zeta p s} \, d_{\zeta p s} \right)
\right.
\np
\left.
         B\hspace{1pt}'_{tp} 
  \, \left( b^\dagger_{\omega p s} \, b_{\omega p s} 
          + d^\dagger_{\omega p s} \, d_{\omega p s} \right) 
\right.
\np
\left.
        C\hspace{1pt}'_{tp} \,
 \left( b_{\zeta  p s}^\dagger \, b_{\omega p s}    
      + b_{\omega p s}^\dagger \, b_{\zeta  p s}
\right.
\right.
\np
\left.
\left.
        d_{\zeta  p s}^\dagger \, d_{\omega p s}  
      + d_{\omega p s}^\dagger \, d_{\zeta  p s} \right)
\right] \\
\es
-
\sum_{~~ps} \hspace{-12pt}\int \, 
        C_{tp} \, p^{+ \, 2}
        ( p_\mu^- - p_\nu^- ) \, 
        ( A_{tp}  - B_{tp}  ) \, 
\nt
     \left( b_{\zeta  p s}^\dagger \, b_{\omega p s}   
          + d_{\omega p s}^\dagger \, d_{\zeta  p s} 
          + b_{\omega p s}^\dagger \, b_{\zeta  p s}   
          + d_{\zeta  p s}^\dagger \, d_{\omega p s} \right)
\np
\sum_{~~ps} \hspace{-12pt}\int \, 
2 \, C^{\,2}_{tp} \, p^{+ \, 2}(p_\mu^- - p_\nu^- ) \, 
\nt
\left[
  b_{\zeta  p s}^\dagger \, b_{\zeta  p s}
+ d_{\zeta  p s}^\dagger \, d_{\zeta  p s} 
- b_{\omega p s}^\dagger \, b_{\omega p s} 
- d_{\omega p s}^\dagger \, d_{\omega p s}
\right] \, .
\nn
\eeq
Equating coefficients in front of the same 
operators on both sides of the last equation, 
one arrives at an infinite set of equations;
6 for every momentum mode $p$, i.e., 3 
equations for every choice of $p$ and spin 
$s$. Namely,
\beq
A'_{tp} 
\es
  2 p^{+ \, 2} \, (p^-_\mu - p^-_\nu) \, C^2_{tp}  \, , \\
B\hspace{1pt}'_{tp} 
\es
- 2 p^{+ \, 2} \, (p^-_\mu - p^-_\nu) \, C^2_{tp}  \, , \\
C\hspace{1pt}'_{tp} 
\es 
-   p^{+ \, 2} \, (p^-_\mu - p^-_\nu) \, 
(A_{tp} - B_{tp}) \, C_{tp} \, .
\eeq
The equations are independent of spin. Moreover, 
they can be written using Eqs.~(\ref{At}), 
(\ref{Bt}) and (\ref{Ct}) as
\beq
\label{At1}
\left( { p^{\perp \, 2} + \mu_t^2 \over p^+ } \right)'
\es
  2 p^{+ \, 2} \, \left( 
{ p^{\perp \, 2} + \mu^2 \over p^+ }
- 
{ p^{\perp \, 2} + \nu^2 \over p^+ } 
\right) 
\nt
\left( { m_t^2 \over p^+ } \right)^2 \, , \\
\label{Bt1}
\left( { p^{\perp \, 2} + \nu_t^2 \over p^+ } \right)'
\es
- 2 p^{+ \, 2} \, \left( 
{ p^{\perp \, 2} + \mu^2 \over p^+ }
- 
{ p^{\perp \, 2} + \nu^2 \over p^+ } 
\right) 
\nt 
\left( { m_t^2 \over p^+ } \right)^2 \, , 
\eeq
\beq
\label{Ct1}
\left( {                    m_t^2 \over p^+ } \right)'
\es 
- p^{+ \, 2} \, \left( 
{ p^{\perp \, 2} + \mu^2 \over p^+ } 
- 
{ p^{\perp \, 2} + \nu^2 \over p^+ } 
\right) 
\nt
\left( 
{ p^{\perp \, 2} + \mu_t^2 \over p^+ }
- 
{ p^{\perp \, 2} + \nu_t^2 \over p^+ } 
\right) \, 
\left( { m_t^2 \over p^+ } \right) \, .
\eeq
It is visible that the momentum variables $p^+$ 
and $p^\perp$ drop out from the infinite set of 
equations and every momentum mode $p$ in the FF 
Fourier expansion of quantum fields evolves 
independently of its spin and only to the extent 
that the mass parameters evolve. These parameters 
evolve according to the set of just 3 equations,
which is the same for all momentum modes and 
spins, see Eqs.~(\ref{mut}), (\ref{nut}) and 
(\ref{mt}) below.

\subsubsection{ Evolution of effective mass parameters }
\label{masses}

The RGPEP equations for the mass parameters are 
\beq
\label{mut}
\left( \mu_t^2 \right)'
\es
  2 \, \left( \mu^2 - \nu^2 \right) \, \left( m_t^2 \right)^2 \, , \\
\label{nut}
\left( \nu_t^2 \right)'
\es
- 2 \, \left( \mu^2 - \nu^2 \right) \, \left( m_t^2 \right)^2 \, , \\
\label{mt}
\left( m_t^2 \right)'
\es 
- \left( \mu  ^2 - \nu  ^2 \right) 
\, 
\left( \mu_t^2 - \nu_t^2 \right) \, m_t^2 \, .
\eeq
These equations for fermions are identical 
to Eqs.~(53), (54), and (55) for bosons in 
Ref.~\cite{bosons}, respectively. They would 
have the same solutions for the same initial 
conditions. However, the initial conditions 
for the fermion mass mixing Hamiltonian are 
different from the initial conditions for 
scalar boson mass mixing Hamiltonian. The 
difference originates in the constraints 
that fermions obey and scalar bosons do not. 
Solutions for fermions are discussed in 
Sec.~\ref{solutions}.  

Note that the replacement of the constant free
Hamiltonian $\cH_f$ in Eq.~(\ref{tnpRGPEP}) by 
the part of the Hamiltonian that contains the 
operators $b^\dagger_\zeta b_\zeta + d^\dagger_\zeta
d_\zeta$ and $b^\dagger_\omega b_\omega +
d^\dagger_\omega d_\omega$, which means a change
in the RGPEP generator mentioned below
Eq.~(\ref{pi-A}) in Appendix~\ref{ARGPEP}, yields
a slightly different set of equations,
\beq
\label{mutW}
\left( \mu_t^2 \right)'
\es
  2 \, \left( \mu_t^2 - \nu_t^2 \right) \, \left( m_t^2 \right)^2 \, , \\
\label{nutW}
\left( \nu_t^2 \right)'
\es
- 2 \, \left( \mu_t^2 - \nu_t^2 \right) \, \left( m_t^2 \right)^2 \, , \\
\label{mtW}
\left( m_t^2 \right)'
\es 
- \left( \mu_t^2 - \nu_t^2 \right)^2 \, m_t^2  \, .
\eeq
This set of 3 equations matches the matrix 
Eq.~(A1) in~\cite{bosons} that resembles 
Wegner's equation~\cite{Wegner1,Wegner2,Kehrein} 
for $2 \times 2$ Hamiltonian matrices. Again, 
since these equations are the same as for bosons, 
the only difference between the fermion and boson
solutions comes from the initial conditions that
reflect the presence of constraints for fermions.
Solutions to Eqs.~(\ref{mutW}), (\ref{nutW}) and
(\ref{mtW}) lead to the same results for $t 
\rightarrow \infty$ as solutions to Eqs.~(\ref{mut}), 
(\ref{nut}) and (\ref{mt}) discussed below.

\subsection{ Solutions to the RGPEP equations }
\label{solutions}

Solutions to Eqs.~(\ref{mut}), (\ref{nut}), 
and (\ref{mt}) are derived in Appendix~\ref{Asolutions}.
They differ from the solutions for bosons
with mass mixing~\cite{bosons} due to the
change of initial conditions for mass terms,
\beq
\left[
\begin{array}{cc} \mu^2 &   m^2 \\
                    m^2 & \nu^2
      \end{array} 
\right] 
& \rightarrow &
\left[
\begin{array}{cc} \mu^2 + m^2 & m(\mu+\nu) \\
                  m(\mu+\nu) & \nu^2 + m^2
      \end{array} 
\right] \, .
\eeq
The fermion initial conditions are the square 
of matrix $M$ in Eq.~(\ref{M}). See also 
Eqs.~(\ref{Pminus}) and (\ref{Msquared}) to
recall how $M^2$ emerges due to constraints. 
As a consequence, the parameter $\epsilon$ 
defined by Eq.~(\ref{epsilon}) in the fermion
case replaces the boson parameter $\epsilon = 
\{ 1 + [2m^2/(\mu^2 - \nu^2)]^2 \}^{1/2}$ when 
$m^2$ is replaced by $m(\mu+\nu)$ and the
ratio $2m^2/(\mu^2-\nu^2)$ becomes $2m/(\mu-\nu)$.

Solutions of the RGPEP equations yield the
diagonal form of $M^2$ when $t \rightarrow \infty$
for $\mu > \nu$, see Appendix~\ref{Asolutions}.
The case $\mu = \nu$ is commented on below. The 
eigenvalues of $M^2$ are $m_1^2$ and $m_2^2$, 
where $m_1$ and $m_2$ are the eigenvalues of 
$M$ given in Eq.~(\ref{m12}). Since $M$ is 
hermitian, its eigenvalues are real. This 
means that the eigenvalues of $M^2$ cannot 
be negative no matter how strong the mixing 
parameter $m$ is. This feature distinguishes 
the mass mixing for fermions from mass mixing 
for bosons. The difference is further 
discussed below. 

The eigenvectors of $M^2$ are the same as
eigenvectors of $M$. Therefore, the angle of
rotation $\varphi$ that appears in the
eigenvectors $v_1$ and $v_2$ in Eq.~(\ref{v12}) 
is reproduced in the RGPEP when $t \rightarrow 
\infty$. This is how the RGPEP solves
the mass mixing theory without any need for
re-quantization. 

We also observe that solutions to the RGPEP 
Eqs.~(\ref{mutW}), (\ref{nutW}), and (\ref{mtW}) 
that are obtained using the generator with 
running effective masses, produce the same 
values of masses in the limit $t \rightarrow
\infty$. For finite values of $t$, the 
corresponding angle of rotation $\varphi_t$ 
(see Appendix~\ref{Asolutions}) is different 
but otherwise there is no difference in 
comparison to solutions to Eqs.~(\ref{mut}), 
(\ref{nut}), and (\ref{mt}) with constant 
masses in the generator.

In solving the RGPEP equations, as 
described in Appendix~\ref{Asolutions},
a convenient variable in place of $t$ is $u 
= \delta \mu^4 \, t$, where $\delta \mu^2 = 
\mu^2 -\nu^2 > 0$. If $\mu^2 = \nu^2$, the 
mass parameters do not evolve with $t$, 
irrespective of the initial value of mass-mixing 
parameter $m$. In this special case, one can 
introduce an auxiliary difference between $\mu$ 
and $\nu$ and one can seek solutions in the 
limit of the auxiliary difference going to 0, 
as had already been suggested in Ref.~\cite{bosons}. 
For example, such artificial splitting of degenerated 
fermion masses would have to be introduced in the 
case of local theories with massless fermions,
including theories with chiral symmetry. A 
prominent example of the FF quanta for which 
a small deviation from mass degeneracy is 
involved in defining the parameter $u$ are 
neutrinos~\cite{neutrinos}. 

\subsection{ Spectrum of the theory }
\label{spectrum}

The initial Hamiltonian, $\hat P^-$ in
Eq.~(\ref{P-bd}), is transformed as a 
result of the RGPEP to 
\beq
\hat P^- \es \cU_t \, \cP_t^- \, \cU_t^\dagger \, , 
\eeq
where $\cP_t^-$ is given in Eq. (\ref{Pt})
and $\cU_t$ is taken from Eq.~(\ref{cUt}). 
Thus,
\beq
\label{hatP-t}
\hat P^- 
\es
\sum_{~~ps} \hspace{-12pt}\int \,
  \left[ A_{tp} 
  \, \left( b^\dagger_{t \zeta p s} \, b_{t \zeta p s} 
          + d^\dagger_{t \zeta p s} \, d_{t \zeta p s} \right)
\right.
\np
\left.
         B_{tp}  
  \, \left( b^\dagger_{t \omega p s} \, b_{t \omega p s} 
          + d^\dagger_{t \omega p s} \, d_{t \omega p s} \right) 
\right.
\np
\left.
        C_{tp} 
  \, \left( b_{t \zeta  p s}^\dagger \, b_{t \omega p s}   
          + b_{t \omega p s}^\dagger \, b_{t \zeta  p s}
\right.
\right.
\np
\left.
\left.
            d_{t \zeta  p s}^\dagger \, d_{t \omega p s}  
          + d_{t \omega p s}^\dagger \, d_{t \zeta  p s} \right)
\right] \, ,
\eeq
where the $t$-dependent annihilation operators are
defined by Eqs.~(\ref{btzetap}), (\ref{dtzetap}),
(\ref{btomegap}), and (\ref{dtomegap}). The
corresponding creation operators are defined
through hermitian conjugation. The coefficients
$A_{tp}$, $B_{tp}$, $C_{tp}$ are defined in Eqs.
(\ref{At}), (\ref{Bt}) and (\ref{Ct}), and the
$t$-dependent mass parameters in them are given in
Eqs.~(\ref{mutA}), (\ref{nutA}),
(\ref{deltamutA}), (\ref{mtA}). The RGPEP secures
that the Hamiltonian $\hat P^-$ as an operator
does not depend on $t$ while the creation and
annihilation operators and coefficients of their
products in $\hat P^-$ do depend on $t$, in such a
way that in the limit of $t \rightarrow \infty$
the mass mixing term disappears, $\lim_{t
\rightarrow \infty} C_{tp} = 0$. 

The eigenvalues and eigenstates of $\hat P^-$ do
not depend on $t$. One can construct the
eigenstates using creation operators corresponding
to any value of $t$ one chooses. Having chosen
operators for some selected value of $t$, one can
apply them to the bare vacuum state $|0\rangle$
and create a basis in the FF Fock space. The bare
vacuum does not depend on $t$ (it is annihilated
by all annihilation operators, irrespective of the
value of $t$). If one chooses certain $t$ for
creation and annihilation operators and
construction of the Fock-space basis, the easiest
Hamiltonian to work with is the one expressed in
terms of the same operators. 

In principle, one can also work with different 
operators for constructing states and Hamiltonians.
This option involves potentially complex formulae 
that include logarithms and other functions of the 
ratios of corresponding scales in complex theories.
For example, such setup is useful in the description 
of form factors and structure functions of hadrons 
because the external probes may distinguish a 
considerably different scale from the one that is 
most convenient for solving the hadron mass eigenvalue
problem. The scale evolution of the parton distributions
appears in the transformation matrix between the 
effective quanta used in the eigenvalue equation 
and the effective quanta corresponding the external
probe scale~\cite{npRGPEP}.

The wave functions of eigenstates in the basis
constructed at some $t$ depend on $t$. In general,
the larger $t$ the more limited the spread of wave
functions in total invariant masses of constituent
states around the eigenvalue mass squared. In the
fermion mass mixing case, the wave functions are
simple to describe because one knows them
exactly.

\subsubsection{ The limit of $ t \rightarrow \infty $ }
\label{tlimit}

The simplest choice of $t$ to work with is $t 
\rightarrow \infty$, since in this case there 
is no mass mixing, $C_{\infty p} = m^2_\infty/p^+ 
= 0$. Thus, the effective theory with $t= \infty$ 
is a theory of free fermions with masses $m_1$ 
and $m_2$, with a correspondingly simple spectrum.
Namely, in the limit of $t \rightarrow \infty$,  
\beq
\label{hatP-infty}
\hat P^- 
\es
\sum_{~~ps} \hspace{-12pt}\int \,
  \left[ { p^{\perp \, 2} + m_1^2 \over p^+} \,
  \left( b^\dagger_{\infty \zeta  p s} \, b_{\infty \zeta  p s} 
       + d^\dagger_{\infty \zeta  p s} \, d_{\infty \zeta  p s} \right)
\right.
\np
\left.
         { p^{\perp \, 2} + m_2^2 \over p^+} \,
  \left( b^\dagger_{\infty \omega p s} \, b_{\infty \omega p s} 
       + d^\dagger_{\infty \omega p s} \, d_{\infty \omega p s} \right) 
\right],
\eeq
where the operators with subscript $\infty$ are
given in Eqs.~(\ref{btzetap}), (\ref{dtzetap}),
(\ref{btomegap}), and (\ref{dtomegap}) with $t =
\infty$, i.e., with the angle $\varphi_\infty$ 
given by Eq.~(\ref{phiinfty}), matching the
angle $\varphi$ found in Eq.~(\ref{angle}) as
an ingredient of the IF re-quantization procedure
in Sec.~\ref{re-quantization},
\beq
\label{cinfty}
\varphi_\infty 
\es 
\varphi
\rs 
- \arctan{ \sqrt{ \epsilon - 1 \over \epsilon + 1} } \, .
\eeq
Hence, the eigenvalues of the Hamiltonian $\hat P^-$ 
in Eq.~(\ref{hatP-infty}) are free FF energies of 
$n_{\infty 1}$ fermions and anti-fermions of mass 
$m_1$ and $n_{\infty 2}$ fermions and anti-fermions 
of mass $m_2$, each with some momentum components 
$p^+$ and $p^\perp$ and spin $z$-axis projection 
$s$, no more than 1 particle in any state with the 
same momentum and spin (i.e., in agreement with the 
Pauli exclusion principle for effective fermions),
\beq
P^-_{\{ (p_{1i}, s_{1i}), i=1,..., n_{\infty 1}\},
     \{ (p_{2j}, s_{2j}), j=1,..., n_{\infty 2}\}}
\nn
= \sum_{i=1}^{n_{\infty 1}} {p_{1i}^{\perp \, 2} + m_1^2 \over p_{1i}^+} 
+
\sum_{j=1}^{n_{\infty 2}} {p_{2j}^{\perp \, 2} + m_2^2 \over p_{2j}^+} \, .
\eeq
The spectrum is degenerate. The eigenstates can 
be closely identified because the RGPEP provides 
expressions for the corresponding creation
operators. A complete set of eigenstates (not 
normalized) is defined by writing 
\beq
\{n_{\infty 1}\} \es
\{ (p_{1i}, s_{1i}), i=1,..., n_{\infty 1}\} \, , \\
\{n_{\infty 2}\} \es
\{ (p_{2j}, s_{2j}), j=1,..., n_{\infty 2}\} \, , \\
\label{eigenstates}
|\{n_{\infty 1}\}, \{n_{\infty 2}\}\rangle
\es
\prod_{i=1}^{n_{\infty 1}} \left( b^\dagger_{\infty \zeta  p_{1i} s_{1i}}
                                  ~or~
                                  d^\dagger_{\infty \zeta  p_{1i} s_{1i}} \right) 
\nt
\prod_{j=1}^{n_{\infty 2}} \left( b^\dagger_{\infty \omega p_{2j} s_{2j}} 
                                  ~or~
                                  d^\dagger_{\infty \omega p_{2i} s_{2i}} \right) 
|0\rangle \, , \nn
\eeq
where $|0\rangle$ denotes the bare vacuum state.
$|0\rangle$ is annihilated by all annihilation 
operators for all values of $t$ and it can be 
treated as one and the same state for all values 
of $t$. 

The eigenstates in Eq.~(\ref{eigenstates}) can 
also be written as combinations of states created 
from the same vacuum state by products of the 
creation operators in the FF Fourier expansions 
of the fields $\hat \zeta$ and $\hat \omega$ in 
Eqs.~(\ref{zetaqftext}) and (\ref{omegaqftext})
at $t=0$. For states with a large number of 
fermions, a simple eigenstate made of the 
effective particles with $t=\infty$, i.e., physical 
fermions, is a complex mixture of many states made 
of bare particles corresponding to $t=0$. 

\subsubsection{ Strong mass mixing }
\label{strong}

The mass mixing interaction with $|m| > 
\sqrt{\mu \nu}$ causes the smaller one of 
two eigenvalues of mass matrix $M$, $m_2$ 
in Eq.~(\ref{m12}), to be negative. The 
RGPEP equations imply that $m_2^2$ is the 
square of mass of a physical fermion. The 
sign of $m_2$ remains undetermined by the
equations of RGPEP because, due to the FF 
constraints, the Hamiltonian depends only 
on $M^2$. The eigenvectors of $M^2$ are the 
same as eigenvectors of $M$ but there is a 
difference between evolving $M^2$ in the 
RGPEP and diagonalizing $M$ in the IF 
re-quantization. 

For $\epsilon > 1$, which is obtained assuming 
$\mu > \nu$ and $m \neq 0$ in Eq.~(\ref{LIFI}) 
or, equivalently, Eq.~(\ref{cLPsi}), the smaller 
one of two diagonal elements of evolving mass 
matrix squared, denoted by $\nu_t^2$ in 
Eq.~(\ref{Bt}) and Appendix~\ref{Asolutions}, 
decreases monotonically to its lowest value of 
$m_2^2 \geq 0$, never reaching 0 if the eigenvalue 
$m_2 \neq 0$. Thus, the RGPEP approaches physical 
solution for $\nu_\infty^2$ when $t \rightarrow 
\infty$ without ever referring to the sign of 
$\nu_t$. The question of a strong mass mixing 
with $ |m| > \sqrt{ \mu \nu}$ is: Where is the 
information about the sign of the negative 
eigenvalue $m_2$ of the mass matrix $M$ stored? 

Using Eq.~(\ref{qs}) for the unconstrained 
fermion fields (and variable $t = s^4$ 
instead of $s$), one can write
\beq
\label{qs+}
\hat \Psi_{t+} \es \cU_t \, \hat \Psi_{0+} \, \cU_t^\dagger
\, .
\eeq
This relation describes the basic transformation 
of quantum degrees of freedom. The complementary 
field components $\Psi_{t-}$ can be obtained 
from the constraints they obey. The constraints 
involve interactions, which in our case are just 
the mass mixing terms that are {\it linear},
not quadratic in the mass parameters. This is 
where the sign of $m_{2t}$ appears. Namely, 
\beq
\label{projectedMt}
\hat \Psi_{t-} 
\es 
{ 1 \over i \partial^+} \, 
( i \alpha^\perp \partial^\perp + \beta M_t) \hat \Psi_{t+} \, ,
\eeq
where the matrix $M_t$ is a root of $M_t^2$.
If one writes
\beq
M_t^2 
\es
\left[ 
\begin{array}{cc} 
m_{1t} & m_{It} \\
m_{It} & m_{2t}
\end{array}
\right]^2
\rs
\left[ 
\begin{array}{cc} 
\mu_t^2 &   m_t^2 \\
  m_t^2 & \nu_t^2
\end{array}
\right] \, ,
\eeq
the smooth solution for $M_t$ that satisfies 
the initial conditions is given by
\beq
m_{1t} \es {1 \over 2} 
\left( \mu + \nu + { \delta \mu_t^2 \over  \mu + \nu } \right) \, , \\
m_{2t} \es {1 \over 2} 
\left( \mu + \nu - { \delta \mu_t^2 \over  \mu + \nu } \right) \, , \\
m_{It} \es { m_t^2 \over  \mu + \nu } \, ,
\eeq
where $\delta \mu_t^2$ and $m_t^2$ are
given in Eqs.~(\ref{deltamutA}) and (\ref{mtA}).
The mass mixing term has the same sign that 
$m_t^2$ has, $m_{1t}$ monotonically increases 
from the positive initial fermion mass $\mu$ 
to the eigenvalue mass $m_1$, and $m_{2t}$ 
monotonically decreases from the positive 
initial fermion mass $\nu$ to the eigenvalue 
mass $m_2$, which may become negative even 
if $m^2 + \nu_t^2$ decreases monotonically 
from $\nu^2$ to $m_2^2$ and thus never 
approaches 0. Thus, one obtains
\beq
\label{hatPsi+text1}
\hat \Psi_t \es \hat \Psi_{t+} + \hat \Psi_{t-} \\
\es
\label{xizetastrong2}
\left[ \begin{array}{c}  
\hat \psi_{1t} + \beta \, m_{It} \hat \psi_{2t+} \\ 
\hat \psi_{2t} + \beta \, m_{It} \hat \psi_{1t+}\,                          
\end{array}  \right] \, ,
\eeq
where the quantum fields
\beq
\hat \psi_{1t}
\es
\left[ 
\begin{array}{c}         
                          \hat \zeta_t  \\
(i \partial^+)^{-1} \,           
\left( \sigma^2 \partial^1 
     - \sigma^1 \partial^2 
     + m_{1t} \right)      \hat \zeta_t  
\end{array}  \right] \, , \\
\hat \psi_{2t}
\es
\left[ 
\begin{array}{c}          \hat \omega_t \\
(i \partial^+)^{-1} \,           
\left( \sigma^2 \partial^1 
     - \sigma^1 \partial^2 
     + m_{2t} \right)      \hat \omega_t
\end{array}  \right] \, ,
\eeq
both have the FF Fourier expansions at $x^+=0$
of the form
\beq
\label{psilt}
\hat \psi_{lt}(x) 
\es \hat \psi_{lt+}(x) + \hat \psi_{lt-}(x) \nn
\es
\sum_{~~ps} \hspace{-13pt}\int 
\left[  u_{ps}(m_{lt}) \, b_{tl ps}       \, e^{-ipx} 
      + v_{ps}(m_{lt}) \, d_{tlps}^\dagger \, e^{ ipx}  \right] , 
\nn
\eeq
where $l = 1$ refers to $\hat \zeta$, 
      $l = 2$        to $\hat \omega$,
the spinors are 
\beq
\label{umul}
u_{ps}(m_{lt}) \es { 1 \over \sqrt{|p^+|} } \,
\left[ \begin{array}{c} p^+ \\ 
-i \sigma^2 p^1 + i \sigma^1 p^2 + m_{lt} 
\end{array} \right] \, \chi_s \, , 
\nn
\eeq
\beq
\label{vmul}
v_{ps}(m_{lt}) \es { 1 \over \sqrt{|p^+|} }\,
\left[ \begin{array}{c} - p^+ \\ 
 +i \sigma^2 p^1 - i \sigma^1 p^2 + m_{lt} 
\end{array} \right] \, \chi_{-s} \, , 
\nn
\eeq
and the annihilation operators are defined
according to Eqs.~(\ref{btzetap}), 
(\ref{dtzetap}), (\ref{btomegap}), and 
(\ref{dtomegap}). In summary, the quantum 
field operators $\hat \psi_{1t}$ and $\hat 
\psi_{2t}$ handle the effective fermions of 
masses $m_{1t}$ and $m_{2t}$ that interact 
through the mass mixing interaction of 
strength $m_{It}$.

When $m^2 > \mu \nu$, the limit of $t \rightarrow 
\infty$ produces negative $m_{2t}$ in $\hat \Psi_t$ 
and $m_{It} \rightarrow 0$ in the Hamiltonian. On 
the other hand, the sign of the mass term with
respect to the momentum dependent terms in the 
spinors in $\hat \Psi_t$ can be changed by making 
a chiral rotation. In the FF representation of 
$\gamma$-matrices, it is visible that chiral 
rotations turn the spin-up and spin-down
$+$-components of spinor fields by opposite angles and
the terms proportional to mass are turned by the angle
opposite to the terms that are proportional to $p^+$
and $p^\perp$. Rotation by angle $\pi/2$ changes the 
sign of the mass terms with respect to the 
momentum-dependent terms.

It might seem that one could make a chiral rotation of 
$\psi_{2t}$ and restore the positivity of the mass 
term with $m_{2t}$ as soon as $m_{2t}$ changes sign
as a function of $t$. However, as long as the mass 
mixing interaction term with $m_{It}$ is present, 
the chiral rotation influences the interaction with 
$\psi_{1t}$. Only for $t \rightarrow \infty$, when 
$m_{It}$ vanishes, one can chirally rotate the field
$\psi_{2\infty}$ independently of the field
$\psi_{1\infty}$, both fields representing physical 
particles. The strong mass mixing thus produces physical 
fermions that are chirally rotated with respect to the 
fermions one starts from. 

\subsubsection{ Effective fermions }
\label{ef}

The effective quantum field operators $\psi_{1t}$ and
$\psi_{2t}$ are constructed in Eq.~(\ref{psilt}) in
Sec.~\ref{strong} according to a general scheme for
building effective quantum field operators using the
RGPEP, see Appendix~\ref{ARGPEP}. The FF Hamiltonian
does not change as a result of re-writing it in terms
of the effective fermion operators. The constancy of
the Hamiltonian as a whole includes the infinite 
additive constant dropped in the process of normal 
ordering. The constant does not depend on $t$ because 
the range of kinematical momentum variables does not 
depend on the interaction and is the same for all 
values of $t$. However, the mass mixing interaction 
strength $m_{It}$ decreases when $t$ increases, becoming 
0 in the limit $t \rightarrow \infty$, where the 
same Hamiltonian is expressed in terms of the creation
and annihilation operators for physical fermions.

The Hamiltonian expressed in terms of the effective
quantum fermion fields corresponding to $t$ is
characterized by two features. One of them is that the
masses $m_{1t}$ and $m_{2t}$ differ from the physical
masses $m_1$ and $m_2$ for as long as $t$ is kept
finite instead of being sent to $\infty$. The other
feature is that the interaction term strength $m_{It}$
is different from 0 for as long as $t$ is kept finite.
These features of our simple model solution suggest
that it may be appropriate also in more complex models,
where exact solutions are not known, to keep due mass
mixing interactions intact in an effective theory for
as long as the effective theory includes any
interactions that are capable of contributing to the
effective mass mixing parameters. By the same token, 
it may be misleading to interpret a theory in terms 
of the degrees of freedom that correspond to a 
diagonalized mass matrix when the other interactions 
are present. 

When there are more than just two different
species of fermions mixed by mass terms, say $f$
different species corresponding to different
``flavors'' or ``families,'' the RGPEP leads to
equations for matrices of dimension $f \times f$.
Such equations do not have known analytic
solutions for a general choice of initial
conditions but they do have exact numerical
solutions that can be found using computers.

\section{ Conclusion }
\label{Conclusion}

It has been demonstrated in Ref.~\cite{bosons}
that the RGPEP provides a solution to the quantum
theory of two kinds of scalar bosons that interact 
with each other through mass-mixing terms. The 
solution avoided the divergent vacuum problem 
of the type that was for a long time considered 
critical to construction of a relativistic quantum 
theory of particles and fields~\cite{DiracDeadWood}. 
In this article, it is shown that the RGPEP also 
provides a solution to the FF theory of two kinds 
of spin-1/2 fermions that interact through mass-mixing 
terms, avoiding the associated fermion vacuum problem 
as well. 

The differential RGPEP equations for effective
mass parameters in the theories of bosons and
fermions turn out to have identical forms.
However, the same equations lead to qualitatively
different solutions in the fermion and boson
theories when the mass mixing interactions are 
strong. The reason is that the initial conditions 
in these theories are set in different ways. Since 
the RGPEP does not a priori rely on any perturbative
expansion, it can tell us precisely what happens
in the theories with arbitrary strength of the
interactions.

The $2 \times 2$ mass matrix that appears in the
scalar bosons theory as an initial condition for
the RGPEP differential equations in a suitable
operator basis has the form (see Ref.~\cite{bosons})
\beq 
M^2_B \es 
\left[
\begin{array}{cc} \mu^2 & m^2 \\ m^2 & \nu^2
  \end{array} 
\right] \, , 
\eeq 
while the analogous initial condition in the fermion 
theory has the form (see Sec.~\ref{solutions})
\beq 
M^2_F 
\es 
\left[ \begin{array}{cc} \mu^2 + m^2 & (\mu + \nu)m \\ 
                        (\mu + \nu)m & \nu^2 + m^2
         \end{array} \right] \, . 
\eeq 
In both cases, $\mu$ and $\nu$ denote the bare 
masses of initial quanta and $m$ denotes the 
strength of the mass mixing interaction terms. 
Once these initial conditions are set, the RGPEP 
yields exact solutions for the creation and 
annihilation operators of effective particles
and the corresponding masses in the effective 
Hamiltonians as functions of the scale parameter
$t$. The parameter can take any value starting 
at 0 and ending at $\infty$. At the end of the 
RGPEP evolution, when $t \rightarrow \infty$, 
one obtains quantum Hamiltonians expressed in 
terms of the creation and annihilation operators 
for physical particles. The masses squared of
the physical bosons and fermions are the eigenvalues 
of the matrices $M_B^2$ and $M_F^2$, respectively.

The key difference between $M_B^2$ and $M_F^2$ is
that one of the eigenvalues of $M_B^2$ is negative
when $|m|$ is larger than $\sqrt{\mu \nu}$ while
$M_F^2$ does not have negative eigenvalues no
matter how large is $m$. The wrong sign of the
mass squared for bosons causes that the
eigenvalues of $\hat P^-$ are unbounded from
below. The fermion theory qualitatively differs
from the boson theory because the wrong sign of
mass squared never appears in fermion theory. This
is a consequence of the fermion constraint
equations that are specific to the FF of
Hamiltonian dynamics and do not appear in theories
of scalar bosons. These constraints produce the
diagonal terms $m^2$ that prevent the off-diagonal
mass mixing terms $m(\mu+\nu)$ from inducing a
negative mass squared for fermions no matter how
large is $m$ in comparison to $\mu$ and $\nu$ and
what its sign is. In other words, the FF theory of
bosons with mass mixing interactions may collapse
due to tachyon solutions when the interaction is
strong while the FF theory of fermions cannot have
tachyon solutions and cannot so collapse.

One can speculate about what may be found
when the RGPEP is applied to theories with 
fermions that include interactions other than
the mass mixing. The fermion mass may appear not
only quadratically but also linearly in physically
relevant interactions. The terms linear in masses
can be considerably different from the mass mixing
terms in our simple model. For example, fermion
masses appear linearly in the photon-electron
interaction terms in QED and in quark-gluon
interaction terms in QCD. Perhaps the effective
masses of lightest fermion species could change 
sign in the RGPEP due to the interactions if the
latter have sufficient strength (considerably 
greater than in QED). If this happens, the 
interactions that are linear in the lightest 
fermion masses and hence sensitive to their 
signs could go through zero. Thus, the RGPEP
could possibly unveil new features of relevant 
effective theories due to the associated chiral
rotations. One might be even forced to limit the 
range of allowed strengths of interactions. On 
the other hand, the RGPEP solutions for the mass 
mixing in both scalar boson and spin-1/2 fermion 
theories suggest that interactions of weak 
strength in comparison to masses can hardly 
cause harm such as a collapse due to tachyon
solutions. 

Even if the lessons learned in the elementary
models with mass mixing and no other interaction
are insufficient to guess the nature of approximate 
solutions that the RGPEP may produce in complex 
theories, both the boson and fermion examples 
already indicate that the RGPEP is capable of 
helping in studies of quantum field theories. 
Some help is certainly needed in generating 
effective interactions in a FF theory of neutrino 
oscillations~\cite{neutrinos}, including generation 
of the neutrino mass terms. Since the vacuum problem 
in the IF of dynamics is far from being understood, 
the RGPEP is of particular relevance as a method of 
study because it appears prepared to provide new 
information without changing the trivial nature of 
the FF vacuum state. This special feature may 
remain valid even in theories as complex as QCD, 
if the features typically associated with a complex 
vacuum in the IF of dynamics are instead associated 
with a potentially rich structure of the FF effective 
Hamiltonian operators. Such possibility had been 
previously suggested in Ref.~\cite{Wilsonetal}. 
If that guess is right, the challenge for the RGPEP 
is to produce the required counterterms and 
generate new interaction terms in effective 
theories.

\begin{appendix}

\section{ FF representation of $\gamma$-matrices }
\label{gamma}

The popular representation of 
$\gamma$-matrices~\cite{DE} adopted 
in~\cite{IF}, is called below the IF 
representation. The IF representation 
leads to the FF projection matrices 
$\Lambda_\pm = \gamma^0 \gamma^\pm/2$ 
that mix all four components of the 
Dirac spinors. This Appendix defines 
the representation of $\gamma$ matrices, 
called below the FF representation, 
in which
\beq
\label{lpm}
\Lambda_+ \es
\left[ \begin{array}{cc}  1 & 0 \\ 
                          0 & 0 \end{array} \right]
\, , \quad
\Lambda_- \rs
\left[ \begin{array}{cc}  0 & 0 \\ 
                          0 & 1 \end{array}
\right] \, .
\eeq
In the FF representation, the unconstrained
parts of the fermion fields $\psi$, i.e., $\psi_+=
\Lambda_+ \psi$, form the two upper components of 
$\psi$, and the dependent parts, i.e., constrained 
by the FF constraint equations, $\psi_-=\Lambda_- 
\psi$ form the two lower components. Thus, the 
quantum field $\hat \psi_+$ can be constructed using 
only two-component spinor fields as described in 
Appendix~\ref{quantumfields}. The same construction 
is used in the case of quantum fields $\hat \psi_+$ 
and $\hat \phi_+$ in Sec.~\ref{FFquantization}. 

The forms (\ref{lpm}) of $\Lambda_\pm$ do not
fully define a representation of the algebra
$\gamma^\alpha \gamma^\beta + \gamma^\beta \gamma^\alpha = 2
g^{\alpha \beta}$. A slightly different representation
from the FF one described below was introduced
before in the context of FF formulation of QCD in
Ref.~\cite{ZH2components}, Sec. II B, see
also~\cite{Wilsonetal}, Sec. IV A. The possibility
of representing the Dirac fermions with only
two-component spinors when one is not interested
in the discrete symmetry of parity, is discussed
in~\cite{Weinberg}, p. 221. In the FF of
Hamiltonian dynamics, the parity symmetry is
dynamical and thus not fully understood in complex
theories, due to the lack of precise solutions. In
the simple model with interactions limited to the
mass mixing, the constraint equations that force
$\psi_-$ to form a complete spinor field in
combination with $\psi_+$ can be conveniently solved 
using the FF representation of $\gamma$-matrices.
The parity symmetry is then exhibited in the
spectrum of solutions for states of physical
particles.

The IF representation we start from is 
($k, l = 1, 2, 3$)~\cite{IF}
\beq
\label{g0IF}
\gamma^0    \es \left[ \begin{array}{cc} 1 &  0  \\ 
                                         0 & -1  \end{array} \right] \, , 
\quad
\gamma^k    \rs \left[ \begin{array}{cc}  0        & \sigma^k \\ 
                                        - \sigma^k & 0        \end{array} \right] \, , \\
\gamma^5 \es i \gamma^0 \gamma^1 \gamma^2 \gamma^3
         \rs   \left[ \begin{array}{cc} 0 & 1  \\ 
                                        1 & 0  \end{array} \right] \, , 
\\
\gamma^5 \gamma^0 \es \left[ \begin{array}{cc} 0 & -1  \\ 
                                               1 &  0  \end{array} \right] \, ,   
\quad 
\gamma^5 \gamma^k 
\rs 
\left[ \begin{array}{cc} - \sigma^k &  0        \\ 
                           0        &  \sigma^k \end{array} \right] \, , \\
\sigma^{\mu \nu} \es {i \over 2} [ \gamma^\mu,
\gamma^\nu] \, , 
\quad
\sigma^{kl} \rs \epsilon^{klm} 
\left[ \begin{array}{cc}    \sigma^m & 0        \\ 
                            0        & \sigma^m
\end{array} \right] \, , \\
\sigma^{0 k} \es 
\left[ \begin{array}{cc}   0        &  i \sigma^k \\ 
                         i \sigma^k &    0        \end{array} \right] 
\rs i \alpha^k \, , \\
\Lambda_\pm \es \gamma^{0}\gamma^\pm /2
            \rs {1 \over 2} 
\left[ \begin{array}{cc}      1           & \pm \, \sigma^3  \\ 
                          \pm \, \sigma^3 &        1         \end{array} \right] \, .
\eeq
Every other choice for the $\gamma$ matrices can
be obtained~\cite{good} using $\tilde \gamma^\mu =
U^\dagger \gamma^\mu U$ with $U^\dagger = U^{-1}$.
The FF representation is obtained by defining a
special $U$ that provides matrices $\Lambda_\pm$
of the form (\ref{lpm}) and at the same time
transforms IF spinors in a specific way. This way
is identified by performing suitable rotations 
of the conventional elements in spinor basis 
for physical fermions at rest. 

In the IF representation for $\gamma$-matrices, the 
basis for constructing spinors of fermions at rest 
can be chosen in the form
\beq
\label{u0IF}
u_\uparrow \es 
\left[ \begin{array}{c} 1 \\ 
                        0 \\
                        0 \\
                        0 \end{array} \right]
\, , \quad
u_\downarrow \rs 
\left[ \begin{array}{c} 0 \\ 
                        1 \\
                        0 \\
                        0 \end{array} \right] 
\, , \\
\label{v0IF}
v_\uparrow \es C \bar{ 
\left[ \begin{array}{c} 1 \\ 
                        0 \\
                        0 \\
                        0 \end{array} \right] }^T
\, , \quad
v_\downarrow \rs C \bar{ 
\left[ \begin{array}{c} 0 \\ 
                        1 \\
                        0 \\
                        0 \end{array} \right] }^T \, , 
\eeq
where $C = i \gamma^2 \gamma^0$ is the charge conjugation 
matrix with properties $C = -C^{-1} = - C^\dagger = - C^T$. 
The above definition assumes that $v = C \bar u^T = 
i \gamma^2 u^*$. One obtains
\beq
v_\uparrow \es 
\left[ \begin{array}{c} 0 \\ 
                        0 \\
                        0 \\
                        1 \end{array} \right]
\, , \quad
v_\downarrow \rs 
\left[ \begin{array}{c} 0 \\ 
                        0 \\
                       -1 \\
                        0 \end{array} \right] \, .
\eeq
Action of the projection matrix $\Lambda_+$ on these 
IF spinors yields $u_+ = \Lambda_+ u$ and $v_+ = 
\Lambda_+ v$. Adjusting normalization to $u_+^\dagger 
u_+ = v_+^\dagger v_+ = 1$, one obtains
\beq
u_{\uparrow +}\es 
{1 \over \sqrt{2}}
\left[ \begin{array}{c} 1 \\ 
                        0 \\
                        1 \\
                        0 \end{array} \right]
\, , \quad
u_{\downarrow +}\rs 
{1 \over \sqrt{2}}
\left[ \begin{array}{c} 0 \\ 
                        1 \\
                        0 \\
                       -1 \end{array} \right] 
\, , \\
v_{\uparrow +}\es 
{1 \over \sqrt{2}}
\left[ \begin{array}{c} 0 \\ 
                       -1 \\
                        0 \\
                        1 \end{array} \right]
\, , \quad
v_{\downarrow +}\rs 
{1 \over \sqrt{2}}
\left[ \begin{array}{c}-1 \\ 
                        0 \\
                       -1 \\
                        0 \end{array} \right] \, .
\eeq
This result involves only two linearly 
independent spinor basis elements,
\beq
u_{\uparrow   +} \es - v_{\downarrow +} \, , 
\quad 
u_{\downarrow +} \rs - v_{\uparrow   +} \, ,
\eeq
which can be used as the new elements of spinor 
basis that are invariant under action of and 
span the image of $\Lambda_+$. Acting on the 
IF spinors at rest with $\Lambda_-$ yields
\beq
u_{\uparrow -}\es 
{1 \over \sqrt{2}}
\left[ \begin{array}{c} 1 \\ 
                        0 \\
                       -1 \\
                        0 \end{array} \right]
\, , \quad
u_{\downarrow -}\rs 
{1 \over \sqrt{2}}
\left[ \begin{array}{c} 0 \\ 
                        1 \\
                        0 \\
                        1 \end{array} \right] 
\, , \\
v_{\uparrow -}\es 
{1 \over \sqrt{2}}
\left[ \begin{array}{c} 0 \\ 
                        1 \\
                        0 \\
                        1 \end{array} \right]
\, , \quad
v_{\downarrow -}\rs 
{1 \over \sqrt{2}}
\left[ \begin{array}{c} 1 \\ 
                        0 \\
                       -1 \\
                        0 \end{array} \right] \, .
\eeq
This result also involves only two linearly 
independent spinor basis elements,
\beq
u_{\uparrow   -} \es v_{\downarrow -} \, , 
\quad
u_{\downarrow -} \rs v_{\uparrow   -} \, ,
\eeq
which can be used as the two complementary 
new elements of spinor basis which are 
invariant under action of and span the 
image of $\Lambda_-$. The complete new 
spinor basis is called here the FF basis. 
Its elements are linear combinations of 
the canonical basis elements with coefficients 
that form the four columns of the matrix
\beq
\label{matrixS}
S \es {1 \over \sqrt{2}}
\left[ \begin{array}{cc} 1        &  1        \\ 
                         \sigma^3 & -\sigma^3  \end{array} \right] 
\, .
\eeq
Spinors written in terms of their coefficients
in the FF basis are marked with the subscript FF.

If an IF spinor is a superposition of the canonical basis 
elements with coefficients $a_i$, $i = 1,2,3,4$, one 
can write
\beq
u_{IF} 
\es 
\left[ \begin{array}{c} a_1 \\ 
                        a_2 \\
                        a_3 \\
                        a_4 \end{array} \right] 
\rs
{ a_1 + a_3 \over \sqrt{2} }
u_{\uparrow +}
+
{ a_2 - a_4 \over \sqrt{2} }
u_{\downarrow +}
\np
{ a_1 - a_3 \over \sqrt{2} }
u_{\uparrow -}
+
{ a_2 + a_4 \over \sqrt{2} }
u_{\downarrow -}
\, .
\eeq
This means that in the FF basis the spinor components are
\beq
u_{FF} \es S^T u_{IF} \, .
\eeq
Since the matrix $S$ is orthogonal, $S^T = S^{-1}$, 
one has $u_{IF} = S u_{FF}$. Therefore, also
\beq
S \,  \gamma_{FF} u_{FF} \es \gamma_{IF} \, u_{IF}
                         \rs \gamma_{IF} \, S u_{FF} 
\eeq
and 
\beq
\gamma_{FF} \es S^T \, \gamma_{IF} \, S \, .
\eeq
Carrying out the required matrix multiplications, 
one obtains the following FF representation of the
$\gamma$-matrices:
\beq
\label{g0FF}
\gamma^0 
\es 
\left[ \begin{array}{cc} 0 & 1 \\ 
                         1 & 0 \end{array} \right]
\, ,
\quad 
\gamma^3 
\rs 
\left[ \begin{array}{cc} 0 & -1  \\ 
                         1 &  0  \end{array} \right]
\, , \\
\label{g12FF}
\gamma^1
\es 
\left[ \begin{array}{cc} -i \sigma^2 &   0        \\ 
                            0        &  i\sigma^2 \end{array} \right]
\, ,
\quad
\gamma^2
\rs 
\left[ \begin{array}{cc}  i \sigma^1 &   0        \\ 
                            0        & -i\sigma^1 \end{array} \right],
\eeq
with $\gamma^k$ for $k = 1, 2, 3$ obtained from 
\beq
\gamma^k 
\es 
{1 \over 2 }
\left[ \begin{array}{cc}   [  \sigma^k , \sigma^3]   & -  \{ \sigma^k , \sigma^3 \} \\ 
                          \{\ \sigma^k , \sigma^3 \} & -   [ \sigma^k , \sigma^3]     \end{array} \right]
\, .
\eeq
Hence, 
\beq
\alpha^1
\es 
\left[ \begin{array}{cc}    0        &  i\sigma^2 \\ 
                         -i \sigma^2 &   0        \end{array} \right]
\, ,
\quad
\alpha^2
\rs 
\left[ \begin{array}{cc}    0        &  -i\sigma^1 \\ 
                          i \sigma^1 &    0        \end{array} \right],
\eeq
and
\beq
\alpha^3 
\es 
\left[ \begin{array}{cc} 1 &  0  \\ 
                         0 & -1  \end{array} \right]
\, . 
\eeq
One also obtains
\beq
\quad
\gamma^5 
\es 
\left[ \begin{array}{cc} \sigma^3 &  0        \\ 
                         0        & -\sigma^3
\end{array} \right] \, , \,
\gamma^5 \gamma^0 
\rs 
\left[ \begin{array}{cc} 0        &  \sigma^3 \\ 
                        -\sigma^3 &   0       \end{array} \right]
\, , \eeq
\beq
\gamma^5 \gamma^\perp
\es 
\left[ \begin{array}{cc} -\sigma^\perp & 0         \\ 
                          0        & -\sigma^\perp \end{array} \right]
\, , \\
\gamma^5 \gamma^3
\es 
\left[ \begin{array}{cc} 0        &  -\sigma^3 \\ 
                        -\sigma^3 &   0       \end{array} \right]
\, ,
\eeq
and
\beq
\sigma^{0 k} \es i\alpha^k \, , \quad
\sigma^{12}  
\rs
\left[ \begin{array}{cc}  \sigma^3 & 0          \\ 
                          0        & \sigma^3 \end{array} \right]
\, , \\
\sigma^{23}   
\es
\left[ \begin{array}{cc}  0        & \sigma^1  \\ 
                          \sigma^1 & 0         \end{array} \right]
\, , \quad
\sigma^{31}   
\rs
\left[ \begin{array}{cc}  0        & \sigma^2  \\ 
                          \sigma^2 & 0         \end{array} \right]
\, .
\eeq
Eqs. (2.8) in~\cite{ZH2components}, or
(4.6) in~\cite{Wilsonetal} define a 
different representation. For example, 
the FF matrices $\gamma^\pm$ are real 
instead of imaginary and the roles of 
$\sigma^1$ and $\sigma^2$ are changed.

\section{ FF construction of spinors }
\label{spinors}

This Appendix defines the spinors that are useful 
in constructing the FF quantum fields of fermions
and solving constraint equations in the model
with mass mixing interactions. The spinors are 
obtained using the FF little group, which belongs 
to the 2nd class distinguished by Wigner~\cite{Wigner}. 
The little group preserves the null four-vector 
$n$ (up to a scale) that defines the front hyperplane 
in space-time through the condition $ nx = x^+ = 0 $, 
where $x$ denotes the co-ordinates of points in 
space-time. The subgroup of the Poincar\'e group 
that preserves the front hyperplane is also called 
the group of kinematical symmetries of the
FF of dynamics; the group elements do not depend
of interactions. 

The construction of spinors adopted here draws on 
Refs.~\cite{mgr,glazekoriginals2,deuteron}. The 
resulting notation differs slightly from the one 
introduced in Ref.~\cite{lb} in Eq.~(A3), due to 
keeping boost matrices for spinors explicitly the 
same for fermions and anti-fermions and using the 
kinematical variable $k_0^+$ instead of a mass 
parameter, see Eqs.~(\ref{Bu}) and (\ref{Bv}) 
below.

\subsection{ Spinors corresponding to momentum $k_0$ }
\label{k0}

According to Wigner~\cite{Wigner}, quantum states 
of a particle are obtained from one state with 
some specified kinematical momentum $k_0$, by 
applying to the specified state operators that 
represent elements of the Poincar\'e group (we 
do not discuss discrete transformations). In the 
FF of quantum theory, in distinction from the IF 
in which boosts depend on interactions, one can 
use the FF kinematical subgroup of the Poincar\'e 
group to construct a fermion state with arbitrary 
momentum. This means that the FF allows one to 
construct the states of moving fermions irrespective 
of interactions while the IF does not allow for 
such construction. 

Let us introduce two basis states for spin-1/2
fermions with momentum $k_0$ and different spin
projections on the $z$-axis. Let the kinematical 
components of the momentum $k_0$ be $k^+_0 \neq 0 $ 
and $k^\perp_0 = 0$. The component $k_0^-$ is left 
unspecified by the kinematics because one needs 
to know the Hamiltonian $P^-$ to determine if 
there exists a preferred value of $k_0^-$. For 
free fermions of mass $\mu$, the corresponding 
$P^-$ would distinguish $k_0^- = \mu^2/k_0^+$. It 
would also be natural to assume $k_0^+ = \mu$ for 
free fermions at rest with respect to the observer 
who constructs a theory. However, at the level of
defining the quantum fields~\cite{Weinberg} and
before one fully understands implications of the
assumed dynamics, it is useful to keep the
kinematical quantity $k_0^+$ in the notation. Such
notation allows one to separate the kinematical
construction of quantum field operators from
making assumptions about dynamics. 

Let the spinors corresponding to the two selected 
fermion states have the form 
\beq
u_{0s} \es S^T \, \sqrt{2 k^+_0} \,
\left[ \begin{array}{c}  \chi_s \\ 
                         0  \end{array} \right] 
\rs
\label{fs}
\sqrt{k^+_0} \,
\left[ \begin{array}{c}  \chi_s \\ 
                         \chi_s \end{array} \right] 
\, ,
\eeq
where the matrix $S$ is defined in Appendix \ref{gamma}
in Eq.~(\ref{matrixS}) and $\chi_s$ with $s=\pm 1$ is 
the standard two-component Pauli spinor for states 
with spin up or down. Namely,
\beq
\chi_1
\es
\chi_\uparrow 
\rs 
\left[ \begin{array}{c}  1 \\ 
                         0 \end{array} \right]
\, , \quad 
\chi_{-1}
\rs
\chi_\downarrow 
\rs 
\left[ \begin{array}{c}  0 \\ 
                         1 \end{array} \right] 
\, . 
\eeq
This choice is motivated by the physical meaning 
of spinors in the IF representation of the
$\gamma$-matrices considered in Appendix
\ref{gamma}; $\chi_s$ corresponds to 
the spin projection on $z$-axis equal $s\hbar/2$, 
irrespective of the value of $k_0^+$. 

Similarly, the spinors for two selected basis states 
of anti-fermions are assumed to have the form
\beq
\label{af1}
v_{0s} \es S^T \, \sqrt{2 k^+_0} \,
\left[ \begin{array}{c}  0       \\ 
                         \varphi_s \end{array} \right] 
\rs
\sqrt{k^+_0} \,
\left[ \begin{array}{c}  \sigma^3 \varphi_s \\ 
                        -\sigma^3 \varphi_s \end{array} \right] 
\, .
\eeq
In accordance with Appendix \ref{gamma}, 
when one introduces the two-component spinor 
for anti-fermions using charge-conjugation 
matrix $C$, so that
\beq
\varphi_s \es -i \sigma^2 \chi_s \, ,
\eeq
one obtains
\beq
\label{af}
v_{0s} \es 
\sqrt{k^+_0} \,
\left[ \begin{array}{c}  -  \sigma^1 \chi_s \\ 
                            \sigma^1 \chi_s \end{array} \right] 
\, .
\eeq
Spinors of fermions with momenta other than 
$k_0$ are obtained using a spinor representation 
of the FF kinematical symmetries.

\subsection{ Spinors for momenta other than $k_0$ }
\label{pfromk0}

Spinors corresponding to states of fermions with 
momenta other than $k_0$ are obtained by applying 
a spinor representation of the Lorentz transformations
built using the FF kinematical Poincar\'e group 
generators of boosts along $z$-axis, $-J^{+ \, -}/2 
= K^3$,  and the mixed boost-rotations, $J^{+1} = 
K^1 + J^2$ and $J^{+2} = K^2 - J^1$, e.g., see 
Refs.~\cite{LeutwylerStern1,LeutwylerSternOscillator}.
The required spinor transformations correspond to 
the Lorentz subgroup of matrices $L$ of the form
\beq
L(a_+) \, x  
\es
\left[ \begin{array}{cccc} 1/a_+ & 0   & 0 & 0 \\ 
                           0     & a_+ & 0 & 0 \\
                           0     & 0   & 1 & 0 \\ 
                           0     & 0   & 0 & 1 \end{array} \right] \,
\left[ \begin{array}{c}    x^- \\ 
                           x^+ \\
                           x^1 \\ 
                           x^2 \end{array} \right] \, , \\
L(a_\perp) \, x 
\es
\left[ \begin{array}{cccc} 1  & a_\perp^2  & 2a_1 & 2a_2 \\ 
                           0  & 1          & 0    & 0    \\
                           0  & a_1        & 1    & 0    \\ 
                           0  & a_2        & 0    & 1    \end{array} \right] \,
\left[ \begin{array}{c}    x^- \\ 
                           x^+ \\
                           x^1 \\ 
                           x^2 \end{array} \right] \, , \\
a_+ \es k_2^+/k_1^+ \, , 
\quad 
a_\perp \rs (k_2^\perp - k_1^\perp) /k_1^+ \, .
\eeq
A four-vector $k_1 = (k_1^-, k_1^+, k_1^\perp)$
with $k_1^- = (\mu^2 + k_1^{\perp \, 2})/k_1^+$, is
changed by these matrices irrespective of the 
value of the mass parameter $\mu$ to
\beq
L(a_+) \, k_1  
\es  
[(\mu^2 + k_1^{\perp \, 2})/k_2^+, k_2^+, k_1^\perp]
\, , \\ 
L(a_\perp) \, k_1 
\es  
[(\mu^2 + k_2^{\perp \, 2})/k_1^+, k_1^+, k_2^\perp]
\, .
\eeq
To transform spinors, one can use a spinor representation 
of the matrix
\beq
&&
L(a_+) \, L(a_\perp)  \nn
\es
\left[ \begin{array}{cccc} 
1/a^+ & a_\perp^2 /a_+ & 2a_1/a^+ & 2a_2/a_+ \\ 
0           & a_+      & 0        & 0        \\
0           & a_1      & 1        & 0        \\ 
0           & a_2      & 0        & 1        \end{array} \right] \, ,
\eeq
which transforms the momentum four-vectors according to
\beq
L(a_+) \, L(a_\perp) \, k_1 
\es
[(\mu^2 + k_2^{\perp \, 2})/k_2^+, k_2^+, k_2^\perp]
\, ,
\eeq
no matter what the value of $\mu$ is. The required spinor 
matrix is
\beq
\label{B21}
&& B(k_2,k_1)  \\
\es 
{ 1 \over \sqrt{ k^+_2 k^+_1} } \, 
\left[
k^+_2   \Lambda_+ 
+
k_1^+ \Lambda_- 
+ 
(k^\perp_2 - k_1^\perp) \, \alpha^\perp \Lambda_+ \right] \, .
\nnn
\eeq
By checking the relations
\beq
B(k_3,k_2) B(k_2,k_1) \es  B(k_3,k_1)       \, , \\
B(k_2,k_1)            \es [B(k_1,k_2)]^{-1} \, , \\
B(k_1,k_1)            \es  1 \, ,
\eeq
one can verify that such spinor matrices form 
a group, as they must as representatives of 
elements of a subgroup of the Lorentz group.

By replacing $k_1$ and $k_2$ in Eq.~(\ref{B21})
by $k_0$ and $p$, respectively, and using the FF 
representation of $\gamma$-matrices defined in
Appendix~\ref{gamma}, one obtains
\beq
\label{Bkk0}
B(p,k_0) 
=
{ 1 \over \sqrt{ p^+ k^+_0} } \, 
\left[ \begin{array}{cc} p^+                & 0     \\ 
       -i \sigma^2 p^1 
       +i \sigma^1 p^2  &
k_0^+ \end{array} \right] \, .
\eeq
The spinors for fermions of momentum $p$,
irrespective of mass parameters that may be 
associated with them in a Hamiltonian, are
defined as
\beq
\label{Bu}
u_{ps} \es B(p,k_0) u_{0s} \, , \\
\label{Bv}
v_{ps} \es B(p,k_0) v_{0s} \, ,
\eeq
where the spinors corresponding to $k_0$
are given by Eqs.~(\ref{fs}) and (\ref{af}). 
In full detail, 
\beq
\label{uks}
u_{ps}
\es
{ 1 \over \sqrt{p^+}} \, 
\left[ \begin{array}{c}   p^+  \\ 
                         -i \sigma^2 p^1 
                         +i \sigma^1 p^2 
                         +k_0^+ \end{array} \right] \, \chi_s \, , \\
\label{vks}
v_{ps} 
\es 
{ 1 \over \sqrt{p^+} } \, 
\left[ \begin{array}{c}  -p^+  \\ 
                         +i \sigma^2 p^1 
                         -i \sigma^1 p^2 
                         +k_0^+ \end{array} \right] \, \chi_{-s} \, .
\eeq
The spinors satisfy relations
\beq
\sum_s u_{ps} \bar u_{ps} \es p \hspace{-4pt}/ + k_0^+ \, , \\
\sum_s v_{ps} \bar v_{ps} \es p \hspace{-4pt}/ - k_0^+ \, , 
\eeq
where 
\beq
p   \es ( p^-, p^+, p^\perp) \, , \\
p^- \es { p^{\perp \, 2} + k_0^{+ \, 2} \over p^+ } \, .
\eeq
For free fermions, one could immediately assume 
that the kinematical parameter $k_0^+$ equals 
the physical fermion mass that appears in the 
free fermion Hamiltonian. In the presence of 
interactions, it is not known prior to solving 
the theory how the fermion mass parameter that 
appears in the Hamiltonian is related to the 
physical fermion mass. The latter situation is 
exemplified by the model with mass mixing 
interactions that is solved using the RGPEP in 
Sec.~\ref{FFm}. In more complex theories, especially 
in QCD, which is expected to explain confinement of 
color, it is important to distinguish between the 
kinematical quantity $k_0^+$ and any dynamically 
determined concept of a quark mass $\mu$.

\subsection{ Spinors in quantum fields }
\label{quantumfields}

The quantum fermion field  
\beq
\label{psiqf}
\hat \psi(x) \es \left[ 
\begin{array}{c} \hat \zeta(x) \\ 
                 \hat \xi(x) 
\end{array}\right]
\eeq
on the front $x^+=0$ can be kinematically 
composed from their Fourier components
using momentum variables $p^+$ and $p^\perp$.
The FF constraint equations in theories
of physical interest, including the mass 
mixing model, cause that the independent 
fermion degrees of freedom are the Fourier 
components of the field $\hat \psi_+ = \Lambda_+ 
\hat \psi$. The field $\hat \psi_-$ is related 
to the field $\hat \psi_+$ through the constraints. 

Using the representation of $\gamma$-matrices 
introduced in Appendix~\ref{gamma}, one has
\beq
\label{psi+qf}
\hat \psi_+(x) \es 
\left[ \begin{array}{c} \hat \zeta(x) \\
                        0 
\end{array}\right] \, ,
\quad 
\hat \psi_-(x) \rs 
\left[ \begin{array}{c} 0 \\
\hat \xi(x) \end{array}\right] \, .
\eeq
Acting with $\Lambda_+$ on the spinors of 
Eqs.~(\ref{uks}) and (\ref{vks}), one obtains 
\beq
\Lambda_+ u_{ps} 
\es
\sqrt{p^+} \left[ \begin{array}{c} \chi_s \\ 0
\end{array}\right] \, , \\
\Lambda_+ v_{ps} 
\es
\sqrt{p^+} \left[ \begin{array}{c} -\sigma^1 \chi_s \\ 0
\end{array}\right] \, .
\eeq
Using these results, one can write
\beq
\hat \zeta(x) 
\es 
\sum_{~~ps} \hspace{-13pt}\int  \, \sqrt{p^+} \,
\left[  b_{ps}  \, e^{-ipx} 
      - d_{ps}^\dagger \, e^{ ipx} \sigma^1 \right] \, \chi_s  \, , 
\nn
\label{zetaqf}
\eeq
where 
\beq
\label{notationqf}
\sum_{~~ps} \hspace{-13pt}\int 
\es
\sum_{s = \, \pm 1} \int_{-\infty}^{+\infty}  { d^2 p^\perp \over (2\pi)^2} \,
\int_0^{+\infty}  {d p^+ \over 2(2\pi) p^+} 
\eeq
and the operators $b_{ps}$ and $d_{ps}$ 
annihilate fermions and anti-fermions, 
respectively. Note that one could use 
$|p^+|$ instead of $p^+$ visible in 
Eqs.~(\ref{zetaqf}) and (\ref{notationqf}) 
because $p^+>0$ in these equations.

The non-zero canonical anti-commutation relations 
at $x^+=0$ read
\beq
\left\{ \hat \psi_+(x), \hat \psi_+^\dagger(x') \right\}
\es 
\Lambda_+ \left\{ \hat \zeta(x), \hat \zeta^\dagger(x') \right\}
\\ 
\label{crqf}
\es \Lambda_+ \, \delta^3(x - x') \, , \\
\left\{ b_{ps}, b^\dagger_{p's'} \right\}
\es
\left\{ d_{ps}, d^\dagger_{p's'} \right\}
\\
\label{crbdqf}
\es
2p^+ (2\pi)^3 \delta^3(p - p') \, \delta_{s s'} \, .
\eeq
The fields $\hat \psi_-$ depend on interactions
through constraints and generally are not related 
to $\hat \psi_+$ in any simple way. In the case of 
a theory of free fermions of mass $\mu$, the Dirac 
equation 
\beq
\label{DE}
(i /\hspace{-5pt}\partial - \mu)\, \psi \es 0 
\eeq
takes the form
\beq
i \partial^- \psi_+ + i \partial^+ \psi_- 
- 
( i \alpha^\perp \partial^\perp + \beta \mu) (\psi_+ + \psi_-) 
\es 0 
\label{projected}
\nn
\eeq
and
\beq
\label{psiminus}
\psi_- 
\es 
{ 1 \over i \partial^+ } 
( i \alpha^\perp \partial^\perp + \beta \mu) \,
\psi_+ \, .
\eeq
Using Eq.~(\ref{psiqf}), one has 
\beq
\label{xizeta}
\hat \xi 
\es
{ 1 \over i \partial^+ } \,
\left( \sigma^2 \partial^1 - \sigma^1 \partial^2 +
\mu \right) \hat \zeta 
\eeq
and
\beq
\hat \psi(x) 
\es \hat \psi_+(x) + \hat \psi_-(x) \nn
\es
\sum_{~~ps} \hspace{-13pt}\int 
\left[  u_{ps}(\mu) \, b_{ps}         \, e^{-ipx} 
      + v_{ps}(\mu) \, d_{ps}^\dagger \, e^{ ipx}  \right] \, , 
\nn
\eeq
where the spinors are 
\beq
\label{umu}
u_{ps}(\mu) \es { 1 \over \sqrt{|p^+|} } \,
\left[ \begin{array}{c} p^+ \\ 
-i \sigma^2 p^1 + i \sigma^1 p^2 + \mu 
\end{array} \right] \, \chi_s \, , 
\nn
\eeq
\beq
\label{vmu}
v_{ps}(\mu) \es { 1 \over \sqrt{|p^+|} }\,
\left[ \begin{array}{c} - p^+ \\ 
 +i \sigma^2 p^1 - i \sigma^1 p^2 + \mu 
\end{array} \right] \, \chi_{-s} \, , 
\nn
\eeq
and the modulus of $p^+$ is freely inserted
using the condition $p^+ > 0$ in the FF Fourier
expansion of fields. These spinors match the 
ones defined kinematically in Eqs.~(\ref{Bu}) 
and (\ref{Bv}) using the momentum $k_0$ and
transformations $B(p,k_0)$ of Eq.~(\ref{Bkk0}), 
if one sets $k^+_0=\mu$. This result is visible 
by comparing Eqs. (\ref{umu}) and (\ref{vmu}) 
with Eqs.~(\ref{uks}) and (\ref{vks}),
correspondingly.

\subsection{ Spinor matrix elements }
\label{matrixel}

The fermion model with mass mixing involves matrix
elements of the form
\beq
\bar u_1 \Gamma u_2 
=
\bar u_{01} \gamma^0 B^\dagger(p_1, k_{01})
\gamma^0 \, \Gamma B(p_2, k_{02}) \, u_{02} \, .
\eeq
The subscripts 1 and 2 refer to the spin labels
and selected momenta $k_{01}$ and $k_{02}$ for
fermions of the type 1 and 2 in construction of
their states and the corresponding quantum field
operators, respectively. Each of these types can 
be associated with a mass $\mu$ or $\nu$ in the 
Hamiltonian. If a theory contains more types 
of fermions than two, the subscripts 1 and 2 may 
each be associated with any mass in the Hamiltonian. 
Since 
\beq
\gamma^0 B^\dagger(p, k_0) \gamma^0 
\es
[B(p, k_0)]^{-1}
\rs
B(k_0,p) \, ,
\eeq
the matrix elements can be written as
\beq
\bar u_1 \Gamma u_2 
\es 
\bar u_{01} B_\Gamma \, u_{02} \, .
\eeq
where 
\beq
B_\Gamma \es B(k_{01}, p_1 )\Gamma B(p_2, k_{02})\, .
\eeq
It is assumed that both $k_{01}^\perp$ and
$k_{02}^\perp$ are 0 and the only non-zero 
kinematical parameters left are $k_{01}^+$ 
and $k_{02}^+$. 

In the fermion model with its interaction limited
to mass mixing, all terms in the FF Hamiltonian 
density are bilinear in the fields. Therefore, 
the kinematical momentum variables that appear 
in the spinor matrix elements in the FF Hamiltonian 
involve one kinematical momentum $p = p_1 = p_2$. 
This means that the matrix elements that count 
involve only the matrix 
\beq
B_\Gamma \es B(k_{01}, p )\Gamma B(p, k_{02}) \\
\es
{ 1 \over p^+ \sqrt{ k_{01}^+ k^+_{02}} } \, 
\left[
k_{01}^+   \Lambda_+
+
p^+ \Lambda_-
-
p^\perp \, \alpha^\perp \Lambda_+ \right]
\nt
\Gamma \, 
\left[
p^+   \Lambda_+ 
+
k_{02}^+ \Lambda_- 
+ 
p^\perp \, \alpha^\perp
\Lambda_+ \right]  \, .
\eeq
For the matrix element for $\Gamma = \gamma^+ = 2 
\gamma^0 \Lambda_+$, one uses
\beq
B_{\gamma^+} 
\es
{ p^+ \gamma^+ \over \sqrt{ k_{01}^+ k^+_{02}} } \, ,
\eeq
and obtains
\beq
\bar u_1 \gamma^+ u_2 
\es
\bar v_1 \gamma^+ v_2 
\rs
2 p^+ \, \chi_1^\dagger \chi_2 
\\
\es
2 p^+ \, \delta_{s_1 s_2} 
\, .
\eeq
These matrix elements do not depend on the
kinematical parameters $k_{01}$ and $k_{02}$
used in the FF construction of quantum fields
for spin-1/2 fermions.

For comparison, one can observe that the 
matrix element with $\Gamma=1$, relevant 
to chiral symmetry, does depend on the 
details of constructing quantum fields. 
Namely,
\beq
B_1 \es B(k_{01},p) \, B(p, k_{02}) \\
\es
{ 1 \over \sqrt{ k^+_{01} k^+_{02} } } \, 
\left[
k_{02}^+ \Lambda_-
+
k_{01}^+ \Lambda_+ \right]
\, , \\
\bar u_1 u_2 
\es
- \bar v_1 v_2 
\rs
\left( k_{01}^+ + k_{02}^+ \right) \,
\delta_{s_1 s_2} \, .
\eeq
Thus, these matrix elements are sensitive to 
the values of $k_{01}^+$ and $k_{02}^+$ used 
in constructing states and fields. If one 
insists on $k_{01}^+=\mu$ and $k_{02}^+=\nu$, 
the matrix elements equal $\mu + \nu$ for the 
same spin projections on $z$-axis of fermions 
of types 1 and 2 at rest. 

Note that the FF $z$-axis is also the direction of
motion for a fermion with $\perp$ momentum 0 and
$+$ momentum different from its mass. Moreover,
the ratio $r = p^+/\mu$, where $\mu$ denotes the
fermion mass, tells one in which direction the
fermion moves: $r > 1$ corresponds to motion down
and $r < 1$ to motion against the $z$-axis. Hence,
the same projection $s$ denotes different
helicities depending on the ratio $r$. These
observations are included here in order to prevent
a confusion of the spin projection $s$ with just 
one value of helicity irrespective of $p^+$. For 
the same reason, the interpretation of $s$ as
related to helicity depends on the ratio of $k_0^+$ 
to $\mu$. When the latter depends on the dynamics, 
one has to be careful in interpreting $s$ in 
terms of helicity.

\section{ Elements of the RGPEP }
\label{ARGPEP}

Elements of the renormalization group procedure
for effective particles (RGPEP) are summarized
below for completeness of the article, following
notation adopted in Ref.~\cite{bosons} that treats
the mass mixing interactions of bosons in a
non-perturbative way. More generally, the RGPEP
development can be traced back to the invention of
the similarity renormalization group
procedure~\cite{GlazekWilson1,GlazekWilson2} and
to the conception of the operator formalism that
allows one to calculate effective Hamiltonians
without limiting their domain in the Fock space
expansion of quantum states, and including
interactions that involve various numbers of
quanta~\cite{Z,ZAPP}. Compact expressions for the
RGPEP in a perturbative series up to the 4th order 
in interaction are given in Ref.~\cite{PnpRGPEP}. 

Effective particles are introduced through a 
transformation
\beq
\label{qs}
\psi_s \es \cU_s \, \psi_0 \, \cU_s^\dagger \, ,
\eeq
where $\psi_s$ is a quantum field operator built 
from creation and annihilation operators for 
effective particles of size $s$ and $\psi_0$ 
is a corresponding quantum field operator built 
from creation and annihilation operators for 
bare quanta of a local theory. The creation and 
annihilation operators are denoted collectively 
by $q_s$ and $q_0$, respectively. All kinematical 
quantum numbers that label operators $q$ are the 
same on both sides of Eq.~(\ref{qs}). Masses are 
considered dynamical. Interpretation of $s$ as 
size is based on the form factors that limit how 
far off energy shell the interactions can extend. 
The value $s=0$ corresponds to absence of form 
factors. For a finite $s$, the effective Hamiltonian 
is band-diagonal on the energy scale and the band 
width is $\sim 1/s$. 

A canonical Hamiltonian density is built from
fields $\psi_0$. A corresponding Hamiltonian 
is a polynomial $\cH_0(q_0)$ with coefficients 
$c_0$ that are functions of the quantum numbers 
labeling operators $q_0$. Similarly, $\cH_t(q_t)$ 
is defined through its coefficients $c_t$. For
dimensional reasons, it is convenient to use
$t=s^4$. The RGPEP starts with the equality
\beq
\label{cHt}
\cH_t(q_t) \es \cH_0(q_0) \, ,
\eeq
which says that the same dynamics is expressed
in terms of different operators for different
values of $t$. The initial condition being set 
at $t=0$, variation of the coefficients $c_t$ 
with $t$ is described by the equation obtained 
by differentiating both sides of 
\beq
\cH_t(q_0) \es \cU^\dagger_t \, \cH_0(q_0) \, \cU_t \, ,
\eeq
with respect to $t$, obtaining
\beq 
\label{ht1}
\cH'_t(q_0) \es
[ \cG_t(q_0) , \cH_t(q_0) ] \, ,
\eeq 
where $\cG_t = - \cU_t^\dagger \cU'_t$ and 
\beq
\label{Usolution}
\cU_t 
\es 
T \exp{ \left( - \int_0^t d\tau \, \cG_\tau
\right) } \, .
\eeq
$T$ denotes ordering in $\tau$. 

The RGPEP generator is defined by
\beq
\label{cGdef}
\cG_t \es [ \cH_f, \cH_{Pt} ] \, ,
\eeq
where $\cH_f$, called the free Hamiltonian, 
is the part of $\cH_0(q_0)$ that does not 
depend on the coupling constants, 
\beq
\label{cHf} 
\cH_f \es
\sum_i \, p_i^- \, q^\dagger_{0i} q_{0i} \, .
\eeq 
The subscript $i$ denotes particle species 
and their quantum numbers. The FF free 
energy of a particle with mass $m_i$ and 
kinematical momentum components $p_i^+$ 
and $p_i^\perp$ is
\beq
\label{pi-A}
p^-_i \es { p_i^{\perp \, 2} + m_i^2 \over p_i^+} \, .
\eeq
We shall also consider $\cH_f$ equal to the 
entire part of $\cH$ of the type $q_0^\dagger 
q_0$ that includes the effective mass parameters 
$m_i$ that do depend on interactions. The operator 
$\cH_{Pt}$ is defined knowing $\cH_t$, 
\beq
\label{Hstructure} 
\cH_t(q_0) =
\sum_{n=2}^\infty \, 
\sum_{i_1, i_2, ..., i_n} \, c_t(i_1,...,i_n) \, \, q^\dagger_{0i_1}
\cdot \cdot \cdot q_{0i_n} \, ,
\eeq 
to be 
\beq
\label{HPstructure} 
\cH_{Pt}(q_0) \es
\sum_{n=2}^\infty \, 
\sum_{i_1, i_2, ..., i_n} \, c_t(i_1,...,i_n) \, 
\nt
\left( {1 \over
2}\sum_{k=1}^n p_{i_k}^+ \right)^2 \, \, q^\dagger_{0i_1}
\cdot \cdot \cdot q_{0i_n} \, .
\eeq 
Thus, ${\cal H}_{Pt}$ differs from ${\cal H}_t$ 
by multiplication of each and every term in it 
by a square of a total $+$ momentum involved
in a term~\cite{npRGPEP,PnpRGPEP}. The 
multiplication leads to preservation of 7 
kinematical symmetries of the FF dynamics
in the RGPEP. 

In summary, the coefficients $c_t$ of products of
operators $q_t$ in the effective Hamiltonians 
$\cH_t(q_t)$, are solutions of the equation 
\beq 
\label{AtnpRGPEP}
\cH'_t \es
\left[ [ \cH_f, \cH_{Pt} ], 
\cH_t \right] \, ,
\eeq 
where all operators are written as polynomials
in $q_0$ and the initial condition is provided
by a regulated canonical Hamiltonian with 
counterterms. The counterterms are calculated 
using a condition that for finite $t$ the 
coefficients $c_t$ with finite arguments do not 
depend on the regularization parameters used in 
the canonical Hamiltonian~\cite{GlazekWilson1}. 
There is no need for calculating counterterms 
in the fermion mass mixing model because the 
coefficients $c_t$ with finite arguments do not 
depend on regularization in this model. The 
regularization dependence in the model is 
limited to an additive constant in $\cH_t$, 
which drops out from Eq.~(\ref{tnpRGPEP}).

The band-diagonal structure of $\cH_t$ can be seen 
using a projector $R$ on a subspace in the Fock 
space. The projected RGPEP equation for $\cH_R = 
R \, \cH_t R$,
\beq 
\label{narrowR2}
\cH_R' \es
\left[ [\cH_f, \cH_{PR}], \cH_R \right] \, ,
\eeq 
implies for constant $\cH_f$ that
\beq
\left( \sum_{mn} |\cH_{Imn}|^2 \right)'
\es
- 2 \sum_{km} (\cM^2_{km} - \cM^2_{mk})^2
|\cH_{Ikm}|^2  \nn
\label{start6x}
& \le & 0 \, ,
\eeq
where $\cH_I = \cH - \cH_f$, $\cH_{PR} =
R \, \cH_{Pt} R$, $\cM_{km}$ denotes an 
invariant mass of the particles in a state 
labeled $k$ that are connected through the 
interaction $\cH_I$ to the particles in a 
state labeled $m$, and the matrix elements 
$\cH_{mn} = \langle m | \cH | n \rangle$ 
are evaluated in the basis built from 
eigenstates $|m\rangle$ of $\cH_f$. According 
to Eq.~(\ref{start6x}), the sum of moduli 
squared of the interaction Hamiltonian matrix 
elements decreases as $t$ increases until all 
off-diagonal matrix elements of the interaction 
Hamiltonian between states with different free 
invariant masses vanish. The width of the narrow 
invariant-mass band in $\cH_R$ is $s^{-1}$. When 
the masses in $\cH_f$ increase with $t$, they 
reduce the right-hand side of Eq.~(\ref{start6x}) 
to more negative values and thus accelerate 
formation of the band-diagonal structure of 
$\cH_t$.

\section{ Solving the RGPEP equations }
\label{Asolutions}

It is convenient to define new variables 
\beq
\alpha \es \mu_t^2/\delta \mu^2 \, , \\
\beta  \es \nu_t^2/\delta \mu^2 \, , \\
\gamma \es   m_t^2/\delta \mu^2 \, ,
\eeq
in terms of which Eqs.~(\ref{mut}), 
(\ref{nut}), and (\ref{mt}), read
\beq
\label{alpha}
\alpha'   
\es   2 \, \gamma^2 \, , \\
\label{beta}
\beta'    
\es - 2 \, \gamma^2 \, , \\
\label{gamma0}
\gamma'
\es  -      \left( \alpha - \beta \right) \, \gamma \, ,
\eeq
where prime denotes differentiation with 
respect to the dimensionless parameter 
$u = \delta \mu^4 t$ and $\delta \mu^2 = 
\mu^2 - \nu^2$. Solutions can be found using
the same method as in Ref.~\cite{bosons}
because the equations are identical. We 
quote only key details here, for completeness.

Eqs.~(\ref{alpha}) and (\ref{beta}) imply that 
$\alpha + \beta$ is a constant and
\beq
\label{delta}
\delta'   
\es   4 \, \gamma^2 \, , \\
\label{gamma1}
\gamma'
\es  - \delta \, \gamma \, ,
\eeq
where $\delta = \alpha - \beta$. Multiplying 
these equations by $2\delta$ and $2\gamma$, 
respectively, one arrives at 
\beq
\label{delta2}
{\delta^2}'   
\es   8 \, \delta \, \gamma^2 \, , \\
\label{gamma2}
{\gamma^2}'
\es  - 2 \delta \, \gamma^2 \, ,
\eeq
which implies constant 
\beq
\epsilon^2 \es \delta^2 + 4 \gamma^2 \rs \cT^2 - 4 \cD \, ,
\eeq
where $\cD = D/\delta \mu^4$, $\cT = T/\delta \mu^2$, 
$D$ is the determinant and $T$ is the trace of $M^2$.
Eliminating $\gamma^2$ from Eq.~(\ref{delta2}), one
obtains an ordinary differential equation 
\beq
\label{delta3}
\delta' \es   \epsilon^2 - \delta^2 \, .
\eeq
Integrations of Eqs.~(\ref{delta3}) and 
then~(\ref{gamma1}), using initial conditions
of Eqs.~(\ref{mu0}), (\ref{nu0}), and (\ref{m0}), 
produce
\beq
\label{mutA}
\mu_t^2 \es m^2 + {1 \over 2} \, (\mu^2 + \nu^2) + { 1 \over 2} \, \delta \mu^2_t \, , \\
\label{nutA}
\nu_t^2 \es m^2 + {1 \over 2} \, (\mu^2 + \nu^2) - { 1 \over 2} \, \delta \mu^2_t \, , \\
\label{deltamutA}
\delta \mu^2_t  \es     \delta \mu^2 \, 
                      { \cosh { x_t } + \epsilon \sinh { x_t }
                \over
                        \cosh { x_t } + \epsilon^{-1} \sinh { x_t } } \, , \\
\label{mtA}
        m_t^2   \es {  m (\mu+\nu)  \over
                        \cosh { x_t } + \epsilon^{-1} \sinh { x_t } } \, , 
\eeq
where  $x_t =  \epsilon \, \delta \mu^4 \, t$, and
$\epsilon = \sqrt{1 + [2 m/(\mu - \nu)]^2}$ as in 
Eq.~(\ref{epsilon}). 

The RGPEP produces a family of Hamiltonians
$\hat P^-_t(b_{t \zeta}, d_{t \zeta }, b_{t \omega }, 
d_{t \omega})$ for $t \ge 0$, members of which are
 obtained from $\cP^-_t(b_\zeta, d_\zeta, b_\omega, 
d_\omega)$ in Eq.~(\ref{Pt}) by replacing operators 
$q_{ps}$, i.e.,
$ b_{\zeta  p s} $,
$ d_{\zeta  p s} $,
$ b_{\omega p s} $,
$ d_{\omega p s} $
and their hermitian conjugates, by $q_{tps}$,
i.e., 
\beq
b_{t \zeta  p s } \es \cU_t \, b_{\zeta  p s} \, \cU^\dagger_t \, , \\
d_{t \zeta  p s } \es \cU_t \, d_{\zeta  p s} \, \cU^\dagger_t \, , \\
b_{t \omega p s } \es \cU_t \, b_{\omega p s} \, \cU^\dagger_t \, , \\
d_{t \omega p s } \es \cU_t \, d_{\omega p s} \, \cU^\dagger_t 
\eeq
and their conjugates, correspondingly. 
In fact, all members of the entire family
are the same, see Eq.~(\ref{cHt}). The 
operator $\cU_t$ is given by Eq.~(\ref{Usolution}), 
as a solution of
\beq
\label{U'}
\cU'_t 
\es 
- \cU_t \, [ \cP^-_f, \cP^-_{Pt} ] \, .
\eeq
Results of Sec.~\ref{application}, 
in particular Eq.~(\ref{generator})
and the fact that the RGPEP respects
the FF kinematical symmetries, imply 
\beq
\label{GP1}
[ \cP^-_f, \cP^-_{Pt} ] 
\es
 \delta \mu^2 \, m_t^2 \, \cA \, , \\
\cA \es
\int [p] \, 
\left( b_{\zeta  p s}^\dagger \, b_{\omega p s}   
     - b_{\omega p s}^\dagger \, b_{\zeta  p s}  
\right.
\np
 \left.
       d_{\zeta  p s}^\dagger \, d_{\omega p s} 
     - d_{\omega p s}^\dagger \, d_{\zeta  p s} \right) \, ,
\eeq
which is a product of the numerical 
factor that depends on $t$ and a 
constant operator. Therefore,
\beq
\label{cUt}
\cU_t \es \exp{ (\varphi_t \cA )} \, , 
\eeq
where
\beq
\label{ctsolution2}
\varphi_t \es 
          \arctan{ \sqrt{ \epsilon + 1 \over \epsilon - 1} } 
        - \arctan{ e^{x_t} \sqrt{ \epsilon + 1 \over \epsilon - 1} } \, .
\eeq
To evaluate
\beq
q_{tps} 
    \es
e^{   \varphi_t \cA } 
\, q_{ps} \,
e^{ - \varphi_t \cA } 
\eeq
one can use
\beq
{[} \cA , b_{\zeta  p s }{]} \es - b_{\omega p s} \, , \\
{[} \cA , b_{\omega p s }{]} \es   b_{\zeta  p s} \, , \\
{[} \cA , d_{\zeta  p s }{]} \es - d_{\omega p s} \, , \\
{[} \cA , d_{\omega p s }{]} \es   d_{\zeta  p s} \, ,
\eeq
and their hermitian conjugates. 
There exist combinations
\beq
b_{p s} \es b_{\zeta p s} + z \, b_{\omega p s} \, , \\
d_{p s} \es d_{\zeta p s} + z \, d_{\omega p s} \, , 
\eeq 
for which one has
\beq
{[} \cA , b_{p s} {]} \es z \, b_{p s} \, , \\
{[} \cA , d_{p s} {]} \es z \, d_{p s} \, , 
\eeq
if $z^2 = -1$. One can use $ z = \pm \,i$ 
for fermions and anti-fermions equally. 
Denoting 
\beq
q_{ps\pm} \es q_{\zeta p s} \pm i \, q_{\omega p s} \, , 
\eeq
one obtains
\beq
e^{   \varphi_t \cA } 
\, q_{ps\pm} \,
e^{ - \varphi_t \cA } 
\es
e^{ \pm \,i \, \varphi_t} \, q_{ps \pm} \, .
\eeq
Using
\beq
b_{\zeta  p s } \es { 1 \over2} (b_{ps+} + b_{ps-}) \, , \\
d_{\zeta  p s } \es { 1 \over2} (d_{ps+} + d_{ps-}) \, , \\
b_{\omega p s } \es {-i \over2} (b_{ps+} - b_{ps-}) \, , \\
d_{\omega p s } \es {-i \over2} (d_{ps+} - d_{ps-}) \, , 
\eeq
one obtains
\beq
\label{btzetap}
b_{t \zeta  p s} \es  
  \cos{\varphi_t} \, b_{\zeta  p s} 
- \sin{\varphi_t} \, b_{\omega p s} \, , \\
\label{dtzetap}
d_{t \zeta  p s} \es  
  \cos{\varphi_t} \, d_{\zeta p s} 
- \sin{\varphi_t} \, d_{\omega  p s} \, , \\
\label{btomegap}
b_{t \omega p s} \es  
  \sin{\varphi_t} \, b_{\zeta  p s} 
+ \cos{\varphi_t} \, b_{\omega p s} \, , \\
\label{dtomegap}
d_{t \omega p s} \es  
  \sin{\varphi_t} \, d_{\zeta p s} 
+ \cos{\varphi_t} \, d_{\omega  p s} \, .
\eeq
These equations provide explicit 
definitions of annihilation operators 
for effective particles corresponding 
to the RGPEP parameter $t = s^4$. The 
corresponding relations for creation 
operators are obtained by hermitian 
conjugation. The inverse relations read
\beq
\label{btzetapinv}
b_{\zeta  p s} \es  
  \cos{\varphi_t} \, b_{t \zeta  p s} 
+ \sin{\varphi_t} \, b_{t \omega p s} \, , \\
\label{dtzetapinv}
d_{\zeta  p s} \es  
  \cos{\varphi_t} \, d_{t \zeta p s} 
+ \sin{\varphi_t} \, d_{t \omega  p s} \, , \\
\label{btomegapinv}
b_{\omega p s} \es  
- \sin{\varphi_t} \, b_{t \zeta  p s} 
+ \cos{\varphi_t} \, b_{t \omega p s} \, , \\
\label{dtomegapinv}
d_{\omega p s} \es  
- \sin{\varphi_t} \, d_{t \zeta p s} 
+ \cos{\varphi_t} \, d_{t \omega  p s} \, .
\eeq
When $t \rightarrow \infty$, one obtains  
\beq
\mu_\infty^2 \es m_1^2 \, , \\ 
\nu_\infty^2 \es m_2^2 \, , \\
  m_\infty^2 \es 0     \, ,
\eeq
where $m_1$ and $m_2$ are the fermion 
eigenvalue masses of Eq.~(\ref{m12}), 
and 
\beq
\label{phiinfty}
\lim_{t \rightarrow \infty} \varphi_t
\es 
\varphi \, ,
\eeq
where $\varphi$ is the angle found 
in Eq.~(\ref{angle}). Thus the RGPEP 
produces a FF quantum Hamiltonian for 
the two species of free fermions that 
were obtained in Sec.~\ref{re-quantization}
in the IF of a theory at the price of 
re-quantization. No such re-quantization
is required in the RGPEP.

\end{appendix}



\end{document}